\documentclass[%
 reprint,
 superscriptaddress,
 amsmath,amssymb,
 aps,
]{revtex4-2}

\usepackage{graphicx}% Include figure files
\usepackage{dcolumn}% Align table columns on decimal point
\usepackage{bm}% bold math
\usepackage[colorlinks=true]{hyperref} % add hypertext capabilities
\usepackage{ragged2e}
\usepackage{ulem}
\usepackage{caption}
\usepackage{subcaption}
\usepackage{orcidlink}
\usepackage{nameref}

\usepackage{xcolor}

\newcommand{\review}[1]{\textcolor{black}{#1}}

\usepackage{graphicx4printer}
\usetikzlibrary{positioning,arrows.meta,fit, calc, backgrounds}
\usepackage{standalone}


\begin{abstract}
Gravitational-wave interferometers exhibit a wide variety of short-duration non-Gaussian transients, commonly referred to as glitches, that complicate the detection of astrophysical signals, bias parameter estimation, and detector characterisation. Existing machine-learning approaches classify glitch morphologies but do not provide a complete mechanism to segment and extract these disturbances from the strain data. We introduce a wavelet-based, saliency-guided framework for the supervised extraction of transient noise. Candidates are first pre-tagged using Uniform Manifold Approximation and Projection, which is also used as a diagnostic of the learned representations. A traditional learning model operating on Continuous Wavelet Transform spectrograms then identifies relevant time-frequency regions through saliency maps. These saliency patterns are transferred to an invertible multiresolution representation via the Discrete Wavelet Transform, where adaptive coefficient masking enables exact reconstruction of both glitch-only and glitch-suppressed waveforms. We demonstrate effective extraction across several representative glitch families, including ``Whistle'' and ``Scattered-Light'' transients, and show robustness in challenging regimes such as low signal-to-noise events and partially overlapping structures, where classical thresholding or band-limited filtering methods typically fail or introduce leakage. The proposed framework offers an interpretable and computationally efficient approach to transient-noise extraction, establishing a foundation for scalable applications to larger glitch catalogs and future observing runs.
\end{abstract}

\begin{document}

\title{Wavelet-Based Extraction of Transient Noise in Gravitational-Wave Interferometers using a Saliency-Guided Learning Architecture}

\author{Christopher Alléné\orcidlink{0009-0001-3859-5420}}
\affiliation{Research Center for Space Science, Advanced Research Laboratories, Tokyo City University, 1-28-1 Tamazutsumi, Setagaya-ku, Tokyo 158-8557, Japan}
\affiliation{Graduate School of Information and Data Science and Department of Design and Data Science, Tokyo City University, 3-3-1 Ushikubo-Nishi, Tsuzuki-ku, Yokohama, Kanagawa 224-8551, Japan}

\author{Dhruv Kumar\orcidlink{0000-0001-8205-0404}}
\affiliation{Department of Physics, National Institute of Technology Agartala, Tripura 799046, India}

\author{Yusuke Sakai\orcidlink{0000-0001-8810-4813}}
\affiliation{Research Center for Space Science, Advanced Research Laboratories, Tokyo City University, 1-28-1 Tamazutsumi, Setagaya-ku, Tokyo 158-8557, Japan}
\affiliation{Graduate School of Information and Data Science and Department of Design and Data Science, Tokyo City University, 3-3-1 Ushikubo-Nishi, Tsuzuki-ku, Yokohama, Kanagawa 224-8551, Japan}

\author{Marco Meyer-Conde\orcidlink{0000-0003-2230-6310}}
\email{marco@tcu.ac.jp}
\affiliation{Research Center for Space Science, Advanced Research Laboratories, Tokyo City University, 1-28-1 Tamazutsumi, Setagaya-ku, Tokyo 158-8557, Japan}
\affiliation{Graduate School of Information and Data Science and Department of Design and Data Science, Tokyo City University, 3-3-1 Ushikubo-Nishi, Tsuzuki-ku, Yokohama, Kanagawa 224-8551, Japan}

\author{Hirotaka Takahashi\orcidlink{0000-0003-0596-4397}}
\affiliation{Research Center for Space Science, Advanced Research Laboratories, Tokyo City University, 1-28-1 Tamazutsumi, Setagaya-ku, Tokyo 158-8557, Japan}
\affiliation{Graduate School of Information and Data Science and Department of Design and Data Science, Tokyo City University, 3-3-1 Ushikubo-Nishi, Tsuzuki-ku, Yokohama, Kanagawa 224-8551, Japan}
\affiliation{Earthquake Research Institute, The University of Tokyo, 1-1-1 Yayoi, Bunkyo-ku, Tokyo 113-0032, Japan}

\maketitle

\preprint{APS/123-QED}

%\wordcount[Abstract Limitation]{sections/0_abstract.tex}{200}
%\wordcount[Mainbody Limitation]{sections/1_introduction.tex sections/2_method.tex sections/3_results.tex sections/4_discussions.tex sections/5_conclusion.tex}{4500}

% Please note: Abbreviations should be introduced at the first mention in the main text – no abbreviations lists. Suggested structure of main text (not enforced) is provided below.

\section{Introduction
    %\wordcount{sections/1_introduction.tex}{1000}
    }
% %%%%%%%%%%%%%%%%%%%%%%%%%%%%%%%%%%%%% %
% This is the introduction section (I). %
% %%%%%%%%%%%%%%%%%%%%%%%%%%%%%%%%%%%%% %

% DESCRIPTION:
% This subsection presents motivations of the paper
%
% KEYPOINTS:
% [X] GW interferometers (LIGO/Virgo/KAGRA) operate at extreme sensitivity
%       and exhibit numerous short-lived non-Gaussian transients (“glitches”).
% [X] Glitches originate from diverse physical mechanisms 
%       (mechanical resonances, scattered light, control loops, environment) 
%       and interfere with astrophysical signal detection, PE, detchar, online/offline pipelines.

\indent Gravitational-wave interferometers such as LIGO~\cite{LIGO:Advanced:2015}, Virgo~\cite{Virgo:Advanced:2014}, and KAGRA~\cite{KAGRA:Overview:2021} operate at extraordinary \review{high} sensitivity, continuously probing strain variations smaller than a proton radius.
As detector performance improves, the data increasingly reveal a rich and complex landscape of short-lived non-Gaussian transients, commonly referred to as \textit{glitches}.
These disturbances arise from diverse physical mechanisms, including mechanical resonances, scattered light, control-loop saturations, and environmental couplings \review{\cite{Soni_2025}}.

% DESCRIPTION:
% Impact of glitches in GW community.
%
% KEYPOINTS:
% [X] False triggers in low-latency searches,
% [X] Masking of GW signals (e.g., GW170817),
% [X] Biases in parameter estimation,
% [X] Modification of non-stationary noise background used for search statistics.
\noindent The impacts of glitches \review{vary}. In detection pipelines \cite{pycbc:Usman_2016, gstlal:sachdev2019gstlalsearchanalysismethods, MBTA:Allene_2025, spiir:chu2021, cwb:Drago_2021}, they can produce false triggers \review{(e.g. \cite{glitschen_2021, Nitz_2018})}, distort search statistics, or mask %\review{\sout{weaker}}
 astrophysical signals \review{as illustrated by the loud glitch overlapping} GW170817~\cite{LIGO:GW170817:2017}.
In parameter estimation, unmodeled transient features can bias inferred source parameters or broaden posterior distributions \review{(e.g. \cite{Powell_2018:glitch_PE, Ghonge_2024:broadband, Edy_2021})}. Even in the absence of a candidate event, glitches occurring in several repeating patterns may contribute to the non-stationary noise background, affecting detector sensitivity curves and background estimation for compact binary coalescence searches \review{\cite{Capano_2017}}.
Although often lasting only milliseconds to seconds, glitches can mimic or prevent astrophysical waveform detection, disrupt the estimation of detector noise, and bias both real-time searches and offline parameter inference \review{(e.g. \cite{Macas_2022:localize, Nuttall_2018})}. The presence of such transients therefore remains a central challenge for detector characterization and robust astrophysical analysis.
Consequently, identifying, characterizing, and mitigating transient noise is essential in the GW science community.

% DESCRIPTION:
% Review of existing ML work and identification of missing capability.
%
% KEYPOINTS:
% [X] Gravity Spy, CNNs, UMAP clustering, etc.
% [X] Successfully classify glitches but DO NOT isolate the glitch itself.
\noindent A substantial effort has been dedicated to classifying glitches using machine-learning techniques (\review{e.g.} \cite{Bahaadini:GlitchClassification:2018,Sakai:UnsupervisedGlitchClassification:2022}). In particular, the Gravity Spy project~\cite{Zevin:GravitySpy:2017} demonstrated that convolutional neural networks (CNNs) trained on volunteer-labeled spectrograms can recognize a large variety of glitch morphologies. However, most existing methods provide labels, embeddings, or cluster assignments~\cite{McInnes:UMAP:2020,Oshino:GlitchClassificationKagraUMAP:2025}, but they do not produce a reconstruction of the part of the \review{timeseries} containing only the glitch.
Such reconstruction capability would be invaluable for generative modeling, glitch injection studies, transient noise simulations, \review{training and evaluating waveform-forecasting machine learning models on data streams containing controlled glitch injections to improve their robustness against transient noise, and validating the performance of gravitational-wave search pipelines under well-characterized perturbations.}\\

% DESCRIPTION:
% Identify main technical obstacle (non-invertibility of TF representations).
%
% KEYPOINTS:
% [X] Commonly used TF representations (Q-transform, CWT, STFT) 
%       are not invertible in GW-data practice.
% [X] Saliency in TF-plane cannot be mapped back to a time-series glitch waveform.
From a signal-processing perspective, this extraction limitation is closely tied to the time-frequency representations traditionally used for glitch identification, such as the Q-transform~\cite{Robinet:Omicron:2020} or the continuous wavelet transform (CWT)~\cite{Daubechies:CWT:1990}. These transforms provide precise and visually interpretable spectrograms, but are not necessarily invertible in the form typically employed in gravitational-wave data analysis~\cite{Virtuoso:Wavelet:2024}.
As a result, \review{glitch structures (e.g. the V or W shape typical of a Whistle)} identified in these representations cannot be cleanly mapped back to the time domain.

% DESCRIPTION:
% Introduce DWT as the solution mechanism.
%
% KEYPOINTS:
% [X] Discrete wavelet transform (DWT) provides an invertible 
%       multiresolution representation enabling direct coefficient manipulation.
% [X] Wavelet techniques are widely used for denoising, but have not yet 
%       been integrated with manifold clustering (e.g., UMAP) to guide 
%       unsupervised glitch extraction in GW data.
\noindent The Discrete Wavelet Transform (DWT)~\cite{Pati:DWT:1990} offers a performant invertible multiresolution representation in which each decomposition level corresponds to a well-defined frequency band with localized temporal support. Unlike conventional spectrograms, DWT coefficients can be selectively manipulated at each \review{decomposition level} and then exactly inverted to reconstruct a modified time series. This property makes the DWT particularly well-suited for reconstructive tasks such as isolating transients, suppressing narrowband disturbances, or removing localized noise bursts. While time-frequency analysis methods, including wavelet representations, have been used extensively in the GW community for non-stationary signal characterisation and gravitational waveform reconstruction~\cite{Cornish:DWT:2020,Cornish:BayesWave:2021}, as well as in traditional signal and image-processing domains~\cite{Parida:DWT:2017}, their full integration with a scalable machine-learning workflows for the purpose of glitch segmentation and extraction has not been previously demonstrated.\\

% DESCRIPTION:
% Summarize contribution of this work.
%
% KEYPOINTS:
% [X] Introduces a wavelet-based framework for extracting 
%   transient noise directly from GW strain using O1-O3 public datasets.
% [X] Uses convolutional feature extraction + CAM/ScoreCAM saliency + UMAP visualisation.
% [X] Maps saliency onto DWT coefficients and reconstructs:
%     (i) glitch-only, (ii) glitch-free waveforms.
In this work, we introduce a wavelet-based extraction framework designed to isolate short-duration transient noise directly from gravitational-wave strain data, using publicly available data from the GW Open Science Center (GWOSC) spanning the O1, O2, and O3 observing runs~\cite{Abbott:GWTC-1:2019,Abbott:GWTC-2:2021,Abbott:GWTC-2.1:2022,Abbott:GWTC-3:2023}. The identification of salient time-frequency regions is performed using a supervised neural network augmented with class-activation map (CAM) techniques~\cite{Zhou:CAM:2015,Chattopadhay:GradCAMPlusPlus:2018}.
%, and guided by a pre-tagging procedure based on Uniform Manifold Approximation and Projection (UMAP)~\cite{McInnes:UMAP:2020}. Because no hand-labeled glitch classes are required, the method is naturally applicable to unlabeled datasets and scalable to future observing runs, where new glitch morphologies may emerge as distinct UMAP clusters. 

The resulting saliency maps highlight the time-frequency structures that the model identifies as anomalous or distinctive. To enable signal reconstruction, this saliency information is transferred to the discrete wavelet domain by mapping the selected time-frequency bins onto the corresponding wavelet decomposition levels. A mask is then constructed to modulate the associated DWT coefficients, either preserving them to reconstruct the glitch signal alone or suppressing them to obtain glitch-mitigated data. The final output is obtained via the inverse DWT, yielding an exact reconstruction of both the transient disturbance and the background signal in the time domain.

% DESCRIPTION:
% Experimental validation across multiple morphologies.
%
% KEYPOINTS:
% [X] Effective across whistle, scattered-light, other common families,
%       with preservation of morphology and no distortion.
% [X] Robust in complex scenarios:
%     - overlapping events (neighboring/sibling glitches),
%     - low-SNR transients at detectability limit.
\noindent We validate this framework on a limited set of the most common glitch families. On this dataset, the method yields extractions that preserve the complex temporal and spectral structure of the glitch, while inducing only minimal bias in the surrounding data. We further investigate challenging scenarios involving overlapping glitch events, including sequences with closely neighboring glitches, sibling transients exhibiting repeating patterns, and low-amplitude structures that are difficult to isolate using classical thresholding techniques.

% DESCRIPTION:
% High-level significance and impact.
%
% KEYPOINTS:
% [X] interpretable, computationally efficient,
% [X] enables controlled glitch injection, generative learning,
%   background estimation, and improved noise modeling.
\noindent Beyond providing a practical approach to glitch segmentation and extraction, this work introduces an interpretable and computationally efficient framework for transient-noise analysis. The combination of saliency maps and wavelet-domain manipulation offers an intuitive link between machine-learning features and physical signal structures. The resulting glitch reconstructions can serve as controlled inputs for generative studies, for evaluating detector behavior under simulated disturbances, and for improving glitch catalogs that may be used in detector characterisation in a systematic approach.\\

\review{% DESCRIPTION:
% Unique point and different aspect from other method
%
% KEYPOINTS:
% [X] cite gwsubstract and compare to it
% [X] cite BayesWave and compare to it
\noindent Unlike previous algorithm as \texttt{gwsubtract} or \texttt{BayesWave} \cite{Davis_2022, Cornish_2015, Ghonge_2024}, which rely on auxiliary witness channels or Bayesian wavelet-based glitch modeling, respectively, the method in this article introduces a saliency-guided deep-learning approach that identifies glitch-related time–frequency structures in CWT spectrograms and uses this information to adaptively mask DWT coefficients for glitch extraction and waveform reconstruction. Its key novelty is combining learned glitch localization with wavelet-domain subtraction, rather than relying primarily on explicit glitch modeling or auxiliary-channel information.
This approach may offer great adaptability to complex glitch morphologies, although its performance depends on the quality and diversity of the training data.}
\review{In the current proof of concept, the qualitative efficiency to identify and subtract the glitch features will be assessed through the dimensionality-reduction algorithm Uniform Manifold Approximation and Projection (UMAP)~\cite{McInnes:UMAP:2020}.}

% DESCRIPTION:
% Structural overview of the paper. (optional)
%
% KEYPOINTS:
% [X] Section Method : explanation of the method and process through CWT, ML,
%    CAM, DWT 
% [X] Section Results : Presentation of UMAP, multiglitch output with fainted (Leo, DFA)
% [X] Section Discussion : scientific implications, future O3GK/O4 applications,
%   integration with YOLO-like scalable architectures. Limitation on GravitySpy classification as input for training
% [X] Section Conclusion
%   wavelet masking, reconstruction metrics.
The remainder of this paper is organized as follows.
The Methods section details the workflow with the data preprocessing, time-frequency representations, the complete learning architecture, wavelet-based masking procedure, and reconstructed metrics.
The Results section presents the wavelet-based specifications underlying the extraction capability of the framework across different glitch morphologies and evaluates its performance under complex conditions.
The Discussion elaborates the broader implications for GW data analysis and outlines future applications to O3GK \cite{KAGRA:O3GK:2023} and O4 datasets, as well as integration with scalable architectures such as YOLO \cite{Redmon:YOLO:2015}, enabling more efficient data segmentation in its latest version, including a transformer architecture \cite{Vaswani:AttentionTransformer:2023,Zhang:ViT-YOLO:2021}.

\section{Method
    %\wordcount{sections/2_method.tex}{3000}
    }
\begin{figure*}
    \centering
    %\fbox{\rule{0pt}{250pt}\rule{0.95\linewidth}{0pt}}
    \begin{tikzpicture}[
    >=Latex,
    bigblock/.style={
        draw,
        fill=blue!15,
        minimum width=2.5cm,
        minimum height=1.2cm,
        rounded corners=4pt,
        align=center
    },
    smallblock/.style={
        draw,
        fill=gray!15,
        minimum width=1.2cm,
        minimum height=0.8cm,
        rounded corners=2pt,
        align=center,
        font=\small
    },
    roundblock/.style={
        draw,
        fill=gray!15,
        minimum width=1.2cm,
        minimum height=0.8cm,
        rounded corners=10pt,
        align=center,
        font=\small
    },
    img/.style={
        draw,
        fill=gray!20,
        minimum width=1cm,
        minimum height=1cm,
        align=center
    },
    squareimg/.style={
        draw,
        fill=orange!20,
        minimum width=1.5cm,
        minimum height=1.5cm,
        align=center,
        font=\small
    },
    modelcircle/.style={
    draw,
    circle,
    fill=red!20,
    minimum size=10mm,
    align=center,
    font=\small
    }
]

\node[img] (dataset) at (0,0) {Dataset};

\node[img, right=10mm of dataset.east, fill=yellow!20] (golddata) at (0,0) {Golden Dataset};
\draw[->, thick] (dataset) -- (golddata);

\node[smallblock, right=10mm of golddata.east] (preproc1) {Preprocess};
\draw[->, thick] (golddata) -- (preproc1);

\node[smallblock, right=10mm of preproc1.east] (pretag) {Pretagging};
\draw[->, thick] (preproc1) -- (pretag);

\node[bigblock, right=10mm of pretag.east, align=left] (training) {\textbf{(a)} \textbf{Training}\\- Clipping augmentation\\- Dilation augmentation\\- Validation \& Test};
\draw[->, thick] (pretag) -- (training);

\node[smallblock, below=16mm of golddata.center] (preproc2) {Preprocess};
\draw[->, thick] (dataset) |- (preproc2);

\node[bigblock, right=10mm of preproc2, align=left, fill=red!20] (class) {\textbf{(b)} \textbf{Classification}\\-Inference, CAM, Bounding \\Box, Contouring};
\draw[->, thick] (preproc2) -- (class);

\node[roundblock, below=16mm of training.center] (mask) {Mask};
\coordinate (midpoint) at ($(training.center)!0.5!(class.center)$);
\draw[->, thick] (training.south) |- (midpoint) -|(class.north);
\draw[->, thick] (class) -- (mask);

\node[roundblock, below=16mm of preproc2.center] (timeseries) {TimeSeries};
\draw[->, thick] (dataset) |- (timeseries);

\node[bigblock, below=14mm of class.center, align=left, fill=green!20] (substract) {\textbf{(c)} \textbf{Glitch Substraction}\\- DWT, Masking, $\mathrm{DWT}^-1$};
\coordinate (midpoint2) at ($(mask.center)!0.5!(substract.center)$);
\draw[->, thick] (timeseries) -- (substract);
\draw[->, thick] (mask.south) |- (midpoint2) -| (substract.north);

\node[roundblock, below=16mm of mask.center] (output) {Glitch TimeSeries};
\draw[->, thick] (substract) -- (output);

\end{tikzpicture}

    %TC:ignore
    \caption{\justifying Workflow overview for transient-noise analysis using preprocessed gravitational-wave strain data. 
    \textbf{(a)} Preprocessing: After applying appropriate preconditioning on a golden set (whitening and state vector checks being applied), Omicron-triggered strain segments are converted into CWT spectrograms, which serve as inputs to the learning model; 
    \textbf{(b)} Learning \& Segmentation: a \textit{ResNet}-based classifier, produces GradCAM++\cite{Chattopadhay:GradCAMPlusPlus:2018} saliency maps that localize transient features under the targeted augmentation strategy; 
    \textbf{(c)} Extraction: saliency information is transferred to the wavelet domain, enabling selective coefficient masking for glitch segmentation and reconstruction.}
    %TC:endignore
    \label{fig:workflow}
\end{figure*}

\review{In the LIGO–Virgo–KAGRA (LVK) detectors, the data streams are continuously monitored for transient events using Omicron \cite{Robinet:Omicron:2020}, a low-latency trigger-generation pipeline based on the Q-transform. Omicron identifies statistically significant excess power in the time–frequency domain and generates triggers characterized by their central time, frequency, duration, and Signal-to-Noise Ratio (SNR). These triggers can be further interpreted in conjunction with the detector state vectors, which provide information on the detector's operating conditions and data quality. Together, these data products help identify candidate glitches and select the corresponding data segments for further analysis.}
\review{In the following, signal will refer to $s(t)$, the sum of the astrophysical signal (null in our case) and the total noise $n_{tot}(t)$.
This total noise can be split in a continuous component $n(t)$ and a transient component, the glitch.}

In this article, we present a multi-step workflow designed to remove glitches from time-series data.
The whole workflow is schematically represented in Fig.~\ref{fig:workflow}.
The method takes as input a time series containing a glitch \review{identified through an Omicron trigger} and outputs a cleaned time series in which the glitch has been subtracted.
First, \review{a segment of} the \review{data} signal \review{$s(t)$ containing an Omicron trigger} is preprocessed and undergo a transformation into a time–frequency representation (CWT), enabling the glitch and its characteristic features to be clearly identified.
A machine-learning classifier is then applied to the resulting spectrogram. By analyzing the model through a CAM, we identify the regions of the spectrogram that contribute most strongly to the classification and correspond to the glitch features.
The resulting saliency map is subsequently used to construct a mask in the DWT domain, allowing the glitch contribution to be removed. Finally, the inverse transformation is applied to reconstruct the cleaned time series.
In the present section, each step of this workflow is described.

\subsection{Continuous Wavelet Transform representation\label{subsec:cwt}} %\review{\sout{Time-frequency representation}} }

Wavelet-based representations provide a complementary description of transient structures through localized decompositions across multiple scales. The CWT of a signal $x(t)$ is defined as
\begin{equation}
W_x(a,b) = \frac{1}{|a|^\frac{1}{p}}\int_{-\infty}^{+\infty} x(t)\,
\psi^{*}\!\left(\frac{t-b}{a}\right)\,
{\rm d}t,
\end{equation}
where $a$ and $b$ denote the scale and time shift, respectively, and $\psi$ is the chosen mother wavelet. The scale parameter $a$ is inversely related to frequency, enabling flexible time-frequency localization.
Here, the $|a|^{-1/p}$ factor corresponds to the $\mathrm{L}_p$ normalization. The L2 normalization ($|a|^{-\frac{1}{2}}$) favoring the lower frequencies by reducing the higher ones keeping signal energy constant, where a $|a|^{-1}$ factor would correspond to a L1 norm which normalize all frequencies the same way enlightening the signal amplitudes \cite{Lilly_2009, Lilly_2017}. In the following the L2-norm is used, allowing the glitches to be more visible across the large considered bandwidth.

Exact reconstruction from a wavelet representation requires satisfaction of the admissibility condition
\begin{equation}
C_{\psi} = \int_{0}^{\infty} \frac{|\hat{\psi}(f)|^2}{f}\, \mathrm{d}f < \infty.
\end{equation}
Although the CWT is admissible in principle, it represents the signal using a highly redundant set of coefficients and is therefore primarily used for visualization and diagnostic analysis rather than exact reconstruction. In addition, it is computationally demanding, as coefficients must be evaluated over a densely sampled range of scales and time shifts. In contrast, DWT constructed from orthonormal or bi-orthogonal bases satisfy the admissibility condition exactly and admit an explicit inverse transform. Their dyadic multi-resolution structure allows efficient implementations with computational complexity scaling as $\mathcal{O}(N)$ or $\mathcal{O}(N\log N)$.\\

Among the many possible wavelet families, the complex Morlet is the one that optimizes the time and frequency resolutions, minimizing the Gabor-Heisenberg inequality:
\begin{equation}
    \sigma_t\sigma_f\geq\frac{1}{4\pi},
\end{equation}
with $\sigma_t$ the time resolution and $\sigma_f$ the frequency resolution.
The complex Morlet is defined as follows:
\begin{equation}
\psi(t)=\frac{1}{\sqrt{\pi B}}\mathrm{e}^{-\frac{t^2}{B}}\mathrm{e}^{2i\pi Ct},
\end{equation}
with $B$ the bandwidth that defines the resolution and $C$ the central frequency with $\sigma_f=\frac{1}{2\pi\sqrt{2B}}$ and $\sigma_t=\sqrt{B/2}$. 
\review{To achieve a suitable time-frequency localization over the frequency range of interest, the parameter $C$ is set to $8$.
This choice represents a compromise between frequency and temporal localization: smaller values of $C$ lead to broader frequency responses, particularly at high frequencies, whereas larger values result in broader temporal responses, especially at low frequencies.
The analyzed frequency range extends from $1\,\mathrm{Hz}$ to $8192\,\mathrm{Hz}$, covering the full frequency range of the gravitational-wave data and capturing most glitch features in the time-frequency representation (e.g. Whistle extends beyond $2\,{\rm kHz}$).
Also, frequencies below $20\,{\rm Hz}$ are included to develop a program that is adaptable for future analysis reaching lower frequencies, although the analysis covers the entire frequency band, we are aware that the frequency range below $20\,{\rm Hz}$ has not been fully calibrated.
Importantly, the purpose of this work is not to perform a standard gravitational-wave signal analysis, but rather to characterize and extract instrumental glitches, whose time-frequency structures may extend beyond the frequency band typically used for astrophysical searches.}
\review{The parameter $B$ is set to $0.25$ to ensure a concentrated time-frequency representation of the signal. Larger values of $B$ result in a more diffuse energy distribution, potentially making glitch-specific features harder to distinguish, particularly for low-amplitude glitches whose energy is comparable to the background noise level. A more concentrated representation is therefore expected to facilitate glitch classification and enable more precise mask construction.}
%In order to minimize the aliasing and according to the frequency bandwidth analyzed (from $1\,{\rm Hz}$ to $8192\,{\rm Hz}$ in order to cover the full frequency range of the GW data and incorporate most of the glitches features in the representation) the parameter $C$ is set to $8$.
%The $B$ parameter is set to $0.25$ such that the signal once transformed is well localized in time and frequency allowing a better classification of the glitch by the model and more precise masks construction.
%The selection of this two values for the parameters get the CWT spectrum closer to the Q-Transform spectrum.

%\review{\sout{The Q-transform can be interpreted as a specific windowed Fourier representation related to wavelet transforms but does not satisfy an admissibility condition and therefore does not admit a practical inverse without additional regularization \cite{Virtuoso:Wavelet:2024}.}}

In this work, the CWT \review{representation is used as input to the trained machine-learning model to classify glitches}, to build the mask and to inspect reconstructed signals and verify the preservation of transient morphology and bandwidth, particularly when computing UMAP representations.
\review{UMAP is a nonlinear dimensionality-reduction technique that maps high-dimensional data, here the CWT spectrograms, into a low-dimensional space while preserving its local structure.}
Exact reconstruction is instead performed using the DWT, which provides an orthogonal multi-resolution decomposition into approximation and detail coefficients at dyadic scales. For applications requiring explicit phase coherence across \review{DWT decomposition levels}, extensions such as the dual-tree complex wavelet transform (DTCWT)~\cite{Kingsbury:DTCWT:2001} may be employed; however, the standard orthogonal DWT is sufficient for the present analysis.

\subsection{Machine Learning algorithm and glitch classification}
\subsubsection{Learning architecture and Pre-tagging.}

The model used in this study is based on a \textit{ResNet-50} architecture\cite{He:ResNet:2015} with \textit{ImageNet}-pretrained weights\cite{Deng:ImageNet:2009}, imposing a fixed input size of $224\times224$ pixels. The network begins with an initial convolutional block consisting of a 7$\times$7 convolution, batch normalization, ReLU activation, and a 3$\times$3 max-pooling layer. 
\review{It is followed by four residual stages containing (3, 4, 6, and 3) bottleneck blocks, with outputs comprising (256, 512, 1024, and 2048) feature maps, respectively.}
After the final residual stage, a 2D adaptive average-pooling layer reduces each feature map to a single spatial value, yielding a 2048-dimensional representation independent of input size. This representation is processed by a fully connected module composed of a linear transformation from 2048 to 128 units with ReLU activation, and a final fully connected linear layer projecting from 128 to $N$ output classes\review{, here $N=3$ for the three glitch classes we focus on.}

The input time series has a sampling frequency of $16{,}384\,\mathrm{Hz}$. The CWT spans the frequency range from $1\,\mathrm{Hz}$ to $8,192\,\mathrm{Hz}$, and the resulting spectrograms are rescaled in logarithmic on the frequency axis. All spectrograms are resized accordingly and replicated across three channels to match the expected RGB input format.

\subsubsection{Training}

The training dataset consists of a balanced sample of 4,500 Gravity Spy glitches drawn from three representative morphologies: \textit{Whistle}, \textit{Blip}, and \textit{Scattered Light}, \review{randomly picked among O1, O2 and O3. Using the same classification as GravitySpy, the variation through observing runs among a given class (especially on \textit{Scattered Light}~\cite{Soni_2021}) isn't expected to be penalizing to the model. This is further discussed in subsection \ref{subsec:umap}.}
%Model training follows a curriculum-learning strategy in which high SNR examples are presented first, and progressively lower-SNR samples are introduced to improve robustness to faint transients.
%The SNR spanning from \texttt{147.83} to \texttt{7.50}. 
The dataset is split into training, validation, and test subsets with proportions of 70\%, 20\%, and 10\%, respectively.\\

% FIG.7 — Data augmentation schematic (time scaling, Q-scaling).
\begin{figure*}%[ht]
    \centering
    \begin{tikzpicture}[
    node distance=4mm and 6mm,
    block/.style={
        draw,
        rounded corners=3pt,
        align=center,
        minimum height=8mm,
        text width=2.6cm,
        fill=white,
        font=\small
    },
    arrow/.style={-Latex, thick},
    title/.style={font=\bfseries\small}
]

\node[title] (t1) {(a) Training vs Validation};

\node[block, below=of t1.west, anchor=west, yshift=-8mm] (a1) {Spectrogram};
\node[block, right=of a1] (b1) {Random Clipping\\$E \sim \mathcal U(5,25)$};

%%%%%%%%%%%%%%%%%%%%%%%%%%%%%%%%%%%%
%% BRANCHES
%%%%%%%%%%%%%%%%%%%%%%%%%%%%%%%%%%%%

% Training branch (top)
\node[block, right=of b1] (c11) 
{Random Dilation\\$z \sim \mathcal U(1,8)$};

%%%%%%%%%%%%%%%%%%%%%%%%%%%%%%%%%%%%
%% RECOMBINATION
%%%%%%%%%%%%%%%%%%%%%%%%%%%%%%%%%%%%

\node[block, right=of c11] (d11) {Normalisation\\$[0,1]$};
\node[block, right=of d11] (e11) {Transformed\\Spectrogram};

%%%%%%%%%%%%%%%%%%%%%%%%%%%%%%%%%%%%
%% ARROWS
%%%%%%%%%%%%%%%%%%%%%%%%%%%%%%%%%%%%

\draw[arrow] (a1) -- (b1) -- (c11) -- (d11) -- (e11);

%%%%%%%%%%%%%%%%%%%%%%%%%%%%%%%%%%%%
%% CASE 3
%%%%%%%%%%%%%%%%%%%%%%%%%%%%%%%%%%%%

\node[title, below=2.5cm of t1.west, anchor=west] (t3) {(b) Inferences and Test};

% First inference

\node[block, below=of t3.west, anchor=west, yshift=-10mm] (a3) {Spectrogram};
\node[block, above right=-3mm and 5mm of a3] (b3) {Fixed Clippings\\$E=E_{\rm avg}$};

\node[block, right=of b3] (c3) {$4$ Fixed Dilations\\$z=1,2,4,8$};

% bloc empilé
\node[block, right=of c3] (d3) {Normalisation\\$[0,1]$};

\node[block, right=of d3] (e3) {4 Transformed\\Spectrograms};

\draw[arrow] (a3)|-(b3)--(c3)--(d3)--(e3);

% Second inference

\node[block, below right=-3mm and 5mm of a3] (b4) {Fixed Clippings\\$E=3, 8, 13, 18, 23$};
%\draw[arrow] ;

\node[block, right=of b4] (c4) {Fixed Dilation\\$z=z_{\rm opti}$};

% bloc empilé
\node[block, right=of c4] (d4) {Normalisation\\$[0,1]$};

\node[block, right=of d4] (e4) {5 Transformed\\Spectrograms};

\draw[arrow](a3.south) |- (b4)--(c4)--(d4)--(e4);

%%%%%%%%%%%%%%%%%%%%%%%%%%%%%%%%%%%%
%% Background boxes
%%%%%%%%%%%%%%%%%%%%%%%%%%%%%%%%%%%%

% Reference coordinates\input{sections/tikz-schemes/data-augmentation}

\begin{scope}[on background layer]

\coordinate (LeftEdge)  at ($(a1.west)+(-5mm,0)$);
\coordinate (RightEdge) at ($(e11.east)+(5mm,0)$);

\coordinate (Top1) at ($(t1.north)+(0,-1mm)$);
\coordinate (Bot1) at ($(c11.south)+(0,-1mm)$);
\node[draw=blue!50, fill=blue!5, rounded corners=6pt,
      fit=(LeftEdge|-Top1)(RightEdge|-Bot1),
      inner sep=6pt] {};

%\coordinate (Top2) at ($(t2.north)+(0,-1mm)$);
%\coordinate (Bot2) at ($(e2.south)+(0,+1mm)$);
%\node[draw=green!50!black, fill=green!5, rounded corners=6pt,
%      fit=(LeftEdge|-Top2)(RightEdge|-Bot2),
%      inner sep=6pt] {};

\coordinate (Top3) at ($(t3.north)+(0,-1mm)$);
\coordinate (Bot3) at ($(e4.south)+(0,-1mm)$);
\node[draw=red!50!black, fill=red!5, rounded corners=6pt,
      fit=(LeftEdge|-Top3)(RightEdge|-Bot3),
      inner sep=6pt] {};
\end{scope}

\end{tikzpicture}

    %TC:ignore
    \caption{\justifying Schematic representation of the data-augmentation strategy applied during training. \textbf{(a)} Random energy scale for both the training and the validation then random or multiple fixed time-axis scaling for respectively the training and the validation data are introduced as controlled 
    perturbations to enhance robustness to temporal dilation or contraction, and C-value for variability in fainted glitch events. \textbf{(b)} Two path for the first or second inference: 1- One fixed energy scale and four time-axis scaling are used resulting in as many transformed spectrogram and the inference; 2- Five fixed energy scale and one optimal dilation factor are applied to the input spectrogram resulting in 5 spectrograms.}
    %TC:endignore
    
    \label{fig:augmentation}
\end{figure*}

\review{During training, we apply clipping to the spectrogram to cap the maximum value of the energy for each pixel of the image we give to the model.
We vary the clipping threshold of the spectrogram amplitudes by a value randomly drawn from a uniform distribution over the interval [3,23].}
The same sampling strategy is used during validation after each epoch. Varying the clipping threshold in this way increases the model’s sensitivity to faint glitches.
%After training, 
%monitoring with Weights \& Biases\cite{wandb}
%one can shows that the model accuracy remains stable at 97.6\% ± 0.3 percentage points for clipping thresholds of 5, 10, 15, 20, and 25.
%After training, the model accuracy was evaluated on a test set of 3000 events for each clipping threshold value (3, 8, 13, 18 and 23). All configurations yielded accuracies in the range $97.5\%–97.8\%$.
%Given the sample size (3000 events), the statistical uncertainty on the accuracy is approximately $\pm0.3$ percentage points at the 68\% confidence level (assuming binomial statistics). The observed variations across clipping thresholds therefore fall within statistical fluctuations.
%These results indicate that, within the range [3, 23], the choice of clipping threshold does not significantly affect the model’s classification performance.
%This indicates that performance is robust with respect to the chosen clipping value.
To reduce confusion between morphologically similar classes (e.g., short-duration whistles and blips), we apply data augmentation based on random temporal dilation before resizing the inputs to a normalized resolution of $224\times224$ pixels (see Fig. \ref{fig:augmentation}).
Specifically, we randomly rescale the temporal dimension so that the resulting signal duration is uniformly distributed between 0.5 and 4 seconds applying a dilation factor in [1, 8].
%During validation and testing, we evaluate four fixed durations (0.5, 1, 2, and 4 seconds) and retain the prediction with the highest average confidence score.
Both augmentations — (1) temporal dilation and (2) energy clipping — are applied before normalizing the data to the range [0, 1], as illustrated in Fig. \ref{fig:augmentation}.
The training and validation phases follow the same augmentation; the test phase however follows the inference process described below.

\subsubsection{Inference for glitch denoising \label{subsubsec:inference}}

Two distinct inference procedures are employed.
The way the data augmentation apply during the inference phase is describe on Fig.~\ref{fig:augmentation}b.
In the first stage, the objective is to localize the glitch within the input.
A single clipping threshold, fixed at 13, is used while all four predefined durations (0.5, 1, 2, and 4 s) are evaluated.
These augmentations are corresponding to the upper path of Fig.~\ref{fig:augmentation}b.
For each duration, the model outputs a confidence score, and the final confidence level is obtained by averaging the scores over all durations.
Glitch localization is then performed using the Class Activation Map (CAM; see below) associated with the prediction exhibiting the highest confidence level.
Once an approximate glitch location has been identified, an optimal duration can be selected.
A second inference step is subsequently carried out using multiple clipping thresholds (3, 8, 13, 18, and 23).
These augmentations are the ones of the lower path on Fig.~\ref{fig:augmentation}b.
For each clipping threshold, the model produces a confidence score.
The clipping value yielding the highest overall confidence level across all classes is retained.
The final classification is then determined from this selected prediction. 
The corresponding CAM is finally used to extract the bounding box and construct the associated mask.

\subsection{Saliency map extraction\label{subsec:extraction}}
\subsubsection{CAM selection and combination\label{subsubsec:cam}}

Beyond classification, this architecture is extended to localize transient structures by coupling GradCAM++ to generate two-dimensional saliency maps that highlight the regions contributing most strongly to the classification decision.

\review{GradCAM is an explainable AI technique used to visualize which parts of an image a model relies on to make an inference. It computes channel weights by spatially averaging the first-order gradients of the target class score with respect to the feature-map activations. GradCAM++ extends this approach by using higher-order derivatives, including second- and third-order derivatives in its classical formulation, to compute spatially dependent weights.}

\review{It was selected because its spatially dependent weighting scheme is designed to improve localization compared with GradCAM. In particular, it can better highlight multiple instances of the same class and preserve contributions from spatially distinct regions, rather than predominantly emphasizing the most discriminative region. These properties are particularly relevant for localizing glitches, whose morphologies may involve multiple spatially distributed features and components.}
\review{However, the resulting weighting coefficients involve denominators that can become zero or numerically unstable for certain activation patterns, preventing the computation of a valid saliency map. In such cases, we use XGradCAM \cite{xgradcam} as a fallback. Its alternative weighting formulation does not require these higher-order derivatives and avoids this specific failure mechanism. Although XGradCAM is motivated by sensitivity and conservation principles, it does not necessarily provide the same localization behavior as GradCAM++; we therefore retain GradCAM++ as our primary method.}
%The GradCAM++ using higher order derivative may fail to compute a map, as a fallback we choose to use XGradCAM \cite{xgradcam}, the map generated by XGradCAM being the most similar to the ones from GradCAM++. This fallback provided less precise CAM but is more stable and won't fail providing the map.

Saliency maps were extracted from the last convolutional layer of the first, second, and fourth (final) bottleneck blocks. These maps were subsequently combined according to :
\begin{equation}
    {\rm cCAM} = {\rm CAM}_4\times({\rm CAM}_1+{\rm CAM}_2)
\end{equation}
where the subscripts denote the corresponding bottleneck level.
This combination strategy provides a trade-off between spatial resolution and robustness to \review{the surrounding noise $n(t)$}. In particular, the saliency map from the first bottleneck preserves fine spatial details, while the multiplication by the fourth bottleneck saliency map enhances robustness against noise.
Furthermore, combining the first and second bottleneck saliency maps enables the detection of glitch features with lower intensity, reducing the risk that these faint structures are suppressed during the multiplication with the fourth bottleneck saliency map.
In the following the CAM will refer to this combined CAM.

% In addition to GradCAM++, which relies on higher-order derivatives and can become numerically unstable (sometimes failing to produce a valid CAM), we use XGradCAM\cite{xgradcam} as a fallback. XGradCAM produces visualizations that are generally sharper and closer to GradCAM++ than to standard GradCAM, while being more numerically stable. It is therefore used when GradCAM++ fails due to vanishing or unstable higher-order derivatives.

\subsubsection{CAM-based glitch localization}

The CAM obtained is flatten along frequency axis taking the maximum values to estimate an effective glitch duration, which is used to define a refined temporal crop $\Delta t$ of the spectrogram.
From this flattening, to select the crop, we search for the excesses of signal above the average value \review{over the flattened CAM curve}.
Ranking the excesses by \review{there} maximum value, the strongest one is selected as the expected glitch time region on the input image.
However, in the case of very large glitches like some Scattered Lights, the signal enlightened by the CAM is broad as well and do not shows an excess of signal after the flattening. Such behaviors are characterized by an average value of the flattened CAM lower than the median value.
In such cases, the temporal crop is applied on the part of the input image having a flattened cam higher than a given threshold.
% The value of this threshold being set to $0.05$ which corresponds to 5\% of the maximum value of a normalized CAM.
\review{This threshold is chosen to remove regions that do not contribute to the classification which indicates they do not comport features from the glitch. 
Such regions have low CAM values (CAM values are between 0 and 1), therefore a threshold of 0.05 is chosen.}

%The specific case of multiple glitches still remains. To prevent such case of focusing only on the most salient one, a selection of the integrated CAM excess is made.
%Each glitch having an Omicron trigger label, in the case of $n$ trigger in the main glitch time window, the $n$ labels are associated to the $n$ strongest excesses. The temporal crop $\Delta t$ can then be applied around the selected excess. 
%In order to be conservative and to keep most of the features, the border of the crop are enlarge of a margin of 5\% of the image size along the time axis.
\review{The second inference is then performed. This second inference uses only one of the four representations (i.e. the four durations: 0.5, 1, 2, 4 seconds). The selection is such that the time span in the representation is the shortest one including the glitch of span $\Delta t$.}
% The second inference is then performed on the zoomed representation among the four possibles that fits the most to the temporal crop, yielding an updated CAM.
Flattening this refined CAM over time provides a corresponding frequency bandwidth, denoted $\Delta f$. The same procedure as the one for the time crop is applied for this frequency crop.
%The difference being the absence of multiple glitch allowing to take the highest excess and the lower threshold set to a forth of the highest value of the time-integrated CAM.
The resulting bounding box $(\Delta t, \Delta f)$, centered around Omicron GPS time, is subsequently used to generate more precise saliency-derived masks for denoising and glitch extraction.

Additionally, spurious CAM artifacts near image boundaries are mitigated by applying mirror padding prior to inference. The padding width is chosen such that the effective region of interest is reduced from $224\times224$ to an effective $192\times192$ pixels, corresponding to the typical spatial extent of edge-related artifacts observed during the early stages of the analysis. This padding strategy suppresses boundary-induced activations without altering the central content of the spectrogram. After inference, the padding is removed, and the resulting CAM is mapped back consistently onto the corresponding CWT scale distributions.

\subsubsection{Mask construction}

The cCAM is used to reweight the input spectrogram in order to enhance regions relevant to the glitch while suppressing unrelated high-amplitude features. The weighted spectrogram is defined as:
\begin{equation}
    w_{px,ij} = \sqrt{{\rm input}_{ij} \times {\rm CAM}_{ij}}.
\end{equation}
where $w_{px,ij}$ denotes the weighted pixel value at position $(i,j)$, ${\rm input}_{ij}$ is the original spectrogram intensity, and ${\rm CAM}_{ij}$ is the corresponding CAM value. Prior to this operation, the CAM is smoothed using a Gaussian blur to reduce high-frequency artifacts.
The weighted spectrogram is subsequently normalized to the interval $[0,1]$. Isocontours are then computed for threshold values ranging from 0.05 to 0.5 in increments of 0.05. For each threshold, a binary mask is constructed such that pixels inside the contour are assigned a value of 1, and those outside are assigned 0.

To determine the optimal threshold, a combined ranking statistic is defined based on a comparison between the normalized distributions of CAM values inside and outside the mask. Let $mask_{ij}$ denote the binary mask. The CAM values inside and outside the mask are defined as:
\begin{equation}
\begin{cases}
    cam_{{\rm in},ij}={\rm CAM}_{ij}\cdot mask_{ij}, \\
    cam_{{\rm out},ij}=1-{\rm CAM}_{ij}\cdot(1-mask_{ij}).
\end{cases}
\end{equation}
Normalized distributions of these two sets of values are constructed and denoted by $p_{\rm in}$ and $p_{\rm out}$, respectively (left panel of Fig.~\ref{fig:wavelet_reconstruction}c).

A single ranking statistic is then formulated to simultaneously account for two complementary objectives: (i) maximizing the part of features inside the mask which minimizes the interesting features outside the mask, and (ii) minimizing the irrelevant signal inside the mask which maximizes the irrelevant signal outside the mask.
Both aspects are captured through the following quantities:
\begin{equation}
\begin{cases}
    \eta_{min}=\sum_{k=1}^{N_{\rm bins}} \min(p_{{\rm in},k}, p_{{\rm out},k}),\\
    \eta_{diff}=\sum_{k=1}^{N_{\rm bins}} |p_{{\rm in},k} - p_{{\rm out},k}|.
\end{cases}
\end{equation}
The term $\eta_{\rm min}$ measures the degree of overlap between $p_{\rm in}$ and $p_{\rm out}$: large values indicate that the CAM activations inside is larger than outside, which is desired. Conversely, $\eta_{\rm diff}$ quantifies the overall separation between the two distributions: larger values correspond to stronger contrast between relevant and non-relevant regions.

These two contributions are combined into a single ranking statistic defined as:
\begin{equation}
    \eta = \eta_{min} - \eta_{diff}.
\end{equation}
Maximizing $\eta$ therefore favors masks that simultaneously reduce the overlap between the inside and outside CAM distributions and enhance their statistical separation. The optimal threshold is selected as the one that maximizes this combined ranking statistic, yielding the mask that best isolates glitch-related structures in the spectrogram.

Once selected, a quick treatment is applied to the binary mask to remove too small components (less than 10 pixels) or small holes in the mask (less than 10 pixels as well). 
Subsequently, to smooth mask boundaries of the binary mask, a distance-based exponential decay is applied,
\begin{equation}
    p_{ij} \leftarrow \max\left(p_{ij}, e^{-d_{ij}^2/\sigma^2}\right),
\end{equation}
with $\sigma = 3$, followed by a Gaussian blur. The resulting masks are used to reconstruct either glitch-only or glitch-suppressed waveforms via the inverse DWT. Masks are made available in the supplementary section in units of $224\times224$.

\subsection{Wavelet Masking}

% FIG.8: (a) Timeseries with thresholding method (universal threshold example)
%        (b) CWT illustrating fragmented segmentation
\begin{figure*}
    \centering
    \includegraphics[width=\linewidth]{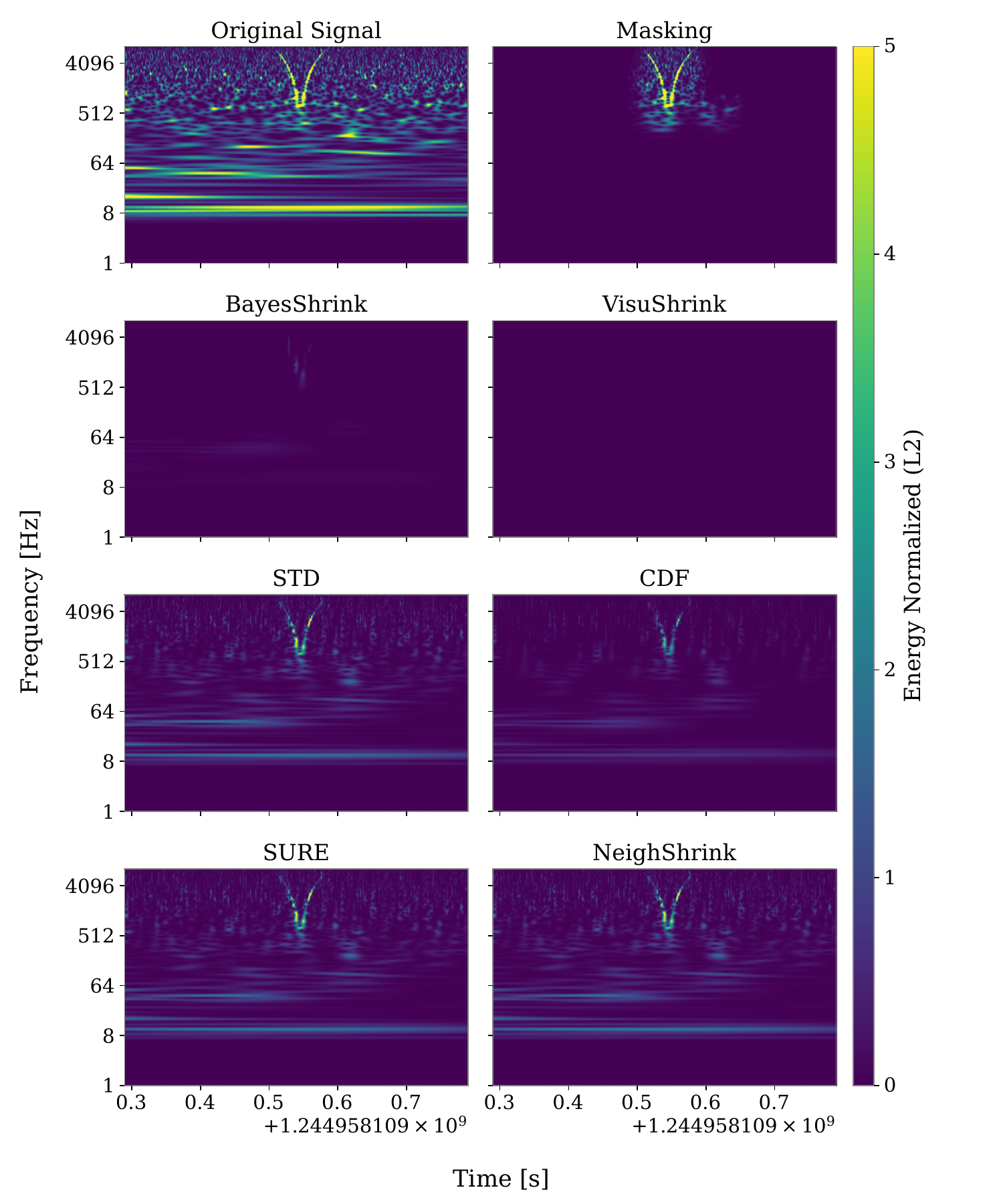}

    %TC:ignore
    \caption{\justifying Example of Whistle glitch (\texttt{LeoGt1PRzA}) subtract by diverse denoising methods applied on DWT. The CWT spectrogram from the denoised TimeSeries are shown and can be compared to the original TimeSeries spectrogram (top left). The Masking plot (top right) follow the method describe in this article. The six other denoising method are originally describe in \cite{bayesshrink} for \texttt{BayesShrink} thresholding, \cite{visushrink} for \texttt{VisuShrink} thresholding, \cite{std_cdf_thresh} for the standard deviation (STD) and Cumulative Distribution Function (CDF) thresholdings, \cite{surethresh} for the \texttt{SURE} thresholding and \cite{neighthresh} for the \texttt{NeighShrink} thresholding. The masking method appear here to be the only one able to isolate the scattered light without the surrounding noise and whistles.}
    %TC:endignore
    \label{fig:threshold_vs_cwt}
\end{figure*}

In this work, the discrete wavelet transform (DWT) provides the reconstruction domain in which saliency-derived masks are applied to selectively preserve or suppress transient contributions. Given a discrete time series $s(t)$ sampled at frequency $f_s$, the DWT decomposes the signal into a hierarchy of approximation and detail coefficients across multiple resolution levels. At the coarsest level $L$, the approximation coefficients capture the low-frequency content of the signal, while the detail coefficients at each level $j$ encode progressively higher-frequency structures with increasing temporal localization. This discrete wavelet transform formula can be written as
\begin{equation}
s(t) = \sum_{k} c_A^{(L)}[k] \, \phi_{L,k}(t)
      + \sum_{j=1}^{L} \sum_{k} c_D^{(j)}[k] \, \psi_{j,k}(t),
\label{eq:dwt}
\end{equation}
where $c_A^{(L)}[k]$ denote the approximation coefficients at the deepest level $L$, $c_D^{(j)}[k]$ are the detail coefficients at scale $j$, and $\phi_{L,k}(t)$ and $\psi_{j,k}(t)$ are the corresponding scaling and wavelet basis functions. Each detail level $j$ is associated with a dyadic frequency band $[f_s/2^{j+1},\, f_s/2^{j}]$
with an effective temporal sampling determined by the downsampling inherent to the transform.

The decomposition depth $L$ is chosen such that the full analysis band from $1\,\mathrm{Hz}$ to $8,192\,\mathrm{Hz}$ is covered. Throughout this work, we employ a Daubechies wavelet of order 8, denoted \texttt{db8}, which provides a suitable compromise between time localization, frequency selectivity, and smoothness for transient glitch reconstruction.

Saliency-based mask obtained in the time-frequency plane is transferred to the DWT domain by constructing a set of one-dimensional windows, one per wavelet band.
For each decomposition level, the mask is cropped to the frequency limits. The maximum values of the mask, indicating to silence (value at one) or not (value at zero) a part of a signal, serve to construct the one-dimensional windows.
These windows are then resampled to match the temporal resolution of the wavelet coefficients.
These windows are resampled to match the temporal resolution of the wavelet coefficients then applied multiplicatively to the detail coefficients prior to DWT inversion recovering timeseries.\\

This approach contrasts with classical wavelet denoising methods based on fixed or universal thresholds~\cite{Donoho:WaveletDenoising:1995}, which often produce fragmented reconstructions and spectral leakage in low–SNR glitch events, as illustrated in Fig.~\ref{fig:threshold_vs_cwt}. 
\review{Wavelet-based thresholding methods share the common principle of separating noise from meaningful signal components by modifying wavelet coefficients according to a selected threshold. Although they all aim to reduce noise while preserving relevant features, they differ in how the threshold is determined and how the coefficients are treated:
\begin{itemize}
    \item \texttt{BayesShrink}: Estimates an adaptive threshold for each wavelet subband using a Bayesian approach and estimates of glitch $g(t)$ and noise $n(t)$ variance;
    \item \texttt{VisuShrink}: Uses a universal threshold based on the noise standard deviation and the number of samples;
    \item \texttt{NeighShrink}: Exploits the local dependency between neighboring wavelet coefficients to adapt the amount of shrinkage;
    \item \texttt{SURE} thresholding: Selects the threshold by minimizing Stein's Unbiased Risk Estimate of the mean squared error;
    \item STD-based thresholding: Determines a threshold from the standard deviation of the wavelet coefficients, often using a multiple of the estimated spread to identify noise or artefact components;
    \item CDF-based thresholding: Uses the cumulative distribution function of the coefficients to determine a threshold associated with a chosen probability or percentile.
\end{itemize}
Overall, the main distinction between these methods lies in their threshold selection strategies and their assumptions about noise and signal structure, which influence the trade-off between noise suppression and preservation of meaningful details.}

In this figure, the thresholding methods either don't extract the targeted glitch like with \texttt{BayesShrink} or \texttt{VisuShrink}, either extract a lot of surrounding noise and partially the targeted glitch like for the other presented methods.
\review{Such thresholding can be efficient to remove the main features of powerful glitches like some Scattered Light of Extremely Loud glitches, but in the cases as the faint Whistle presented in Fig.~\ref{fig:threshold_vs_cwt} in which a structure is present and visible in the noise but whose energy is near the surrounding noise level, the thresholding method based on the statistical properties of the signal more than a structure related search fail to catch and separate the glitch from the rest of the noise.}
\review{This shows also why Gaussianity tests are limited to evaluate subtraction efficiency when such fainted glitches occur.}
By contrast, saliency-guided masking\review{, focusing on the structures identified by a trained model,} selectively preserves wavelet coefficients associated with the glitch morphology while suppressing unrelated noise, especially other glitches nearby.
\review{However, this capability to focus on structures can be limited by the capability of the model to learn and identify for instance when multiple glitches are superimposed.}
\section{Results
    %\wordcount{sections/3_results.tex}{2000}
    }
% %%%%%%%%%%%%%%%%%%%%%%%%%% %
% This is the result section %
% %%%%%%%%%%%%%%%%%%%%%%%%%% %

% 0. Present the Learning curve and the confusion matrix
% 1. Present results for the example of one glitch : OUL...
% 2. Present the results on a set of glitches : UMAP evolution
% 3. Present the results on multiglitch with fainted (Leo, DFA)

\subsection{Training of the model}

% FIG.2: (a) Learning curves (training + validation with/without augmentation)
%        (b) Confusion matrix (classnames, ML efficiency, data counts)
%
\begin{figure*}
    \centering
    \begin{minipage}{0.98\linewidth}
        \centering
        % \fbox{\rule{0pt}{175pt}\rule{0.9\linewidth}{0pt}}
        \includegraphics[width=\linewidth]{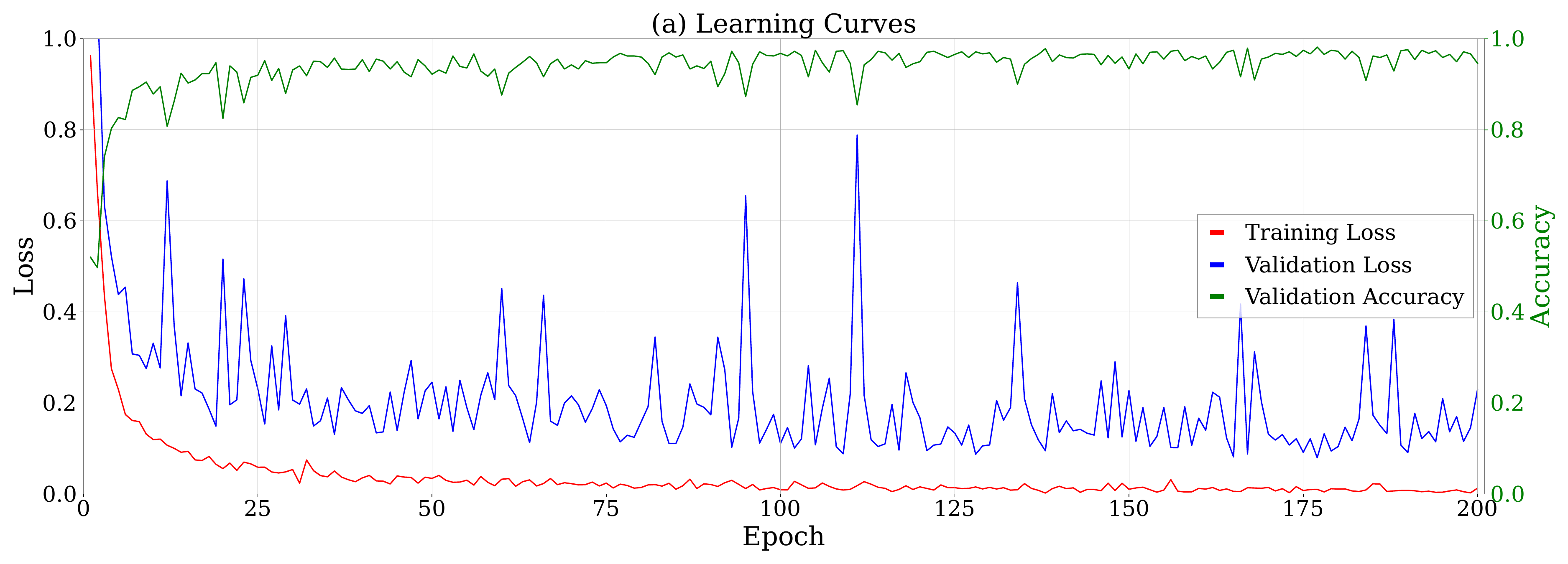}
    \end{minipage}
    \hfill
    \begin{minipage}{0.49\linewidth}
        \centering
        % \fbox{\rule{0pt}{175pt}\rule{0.9\linewidth}{0pt}}
        \includegraphics[width=\linewidth]{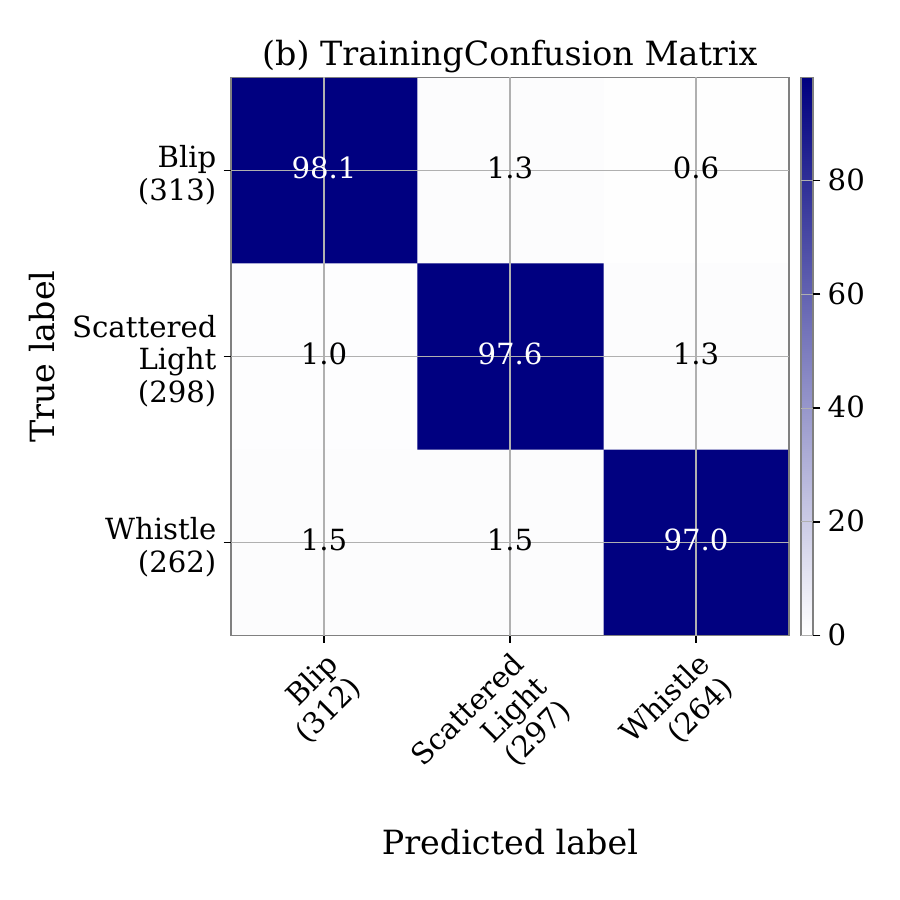}
    \end{minipage}
    \hfill
    \begin{minipage}{0.49\linewidth}
        \centering
        %\fbox{\rule{0pt}{175pt}\rule{0.9\linewidth}{0pt}}
        \includegraphics[width=\linewidth]{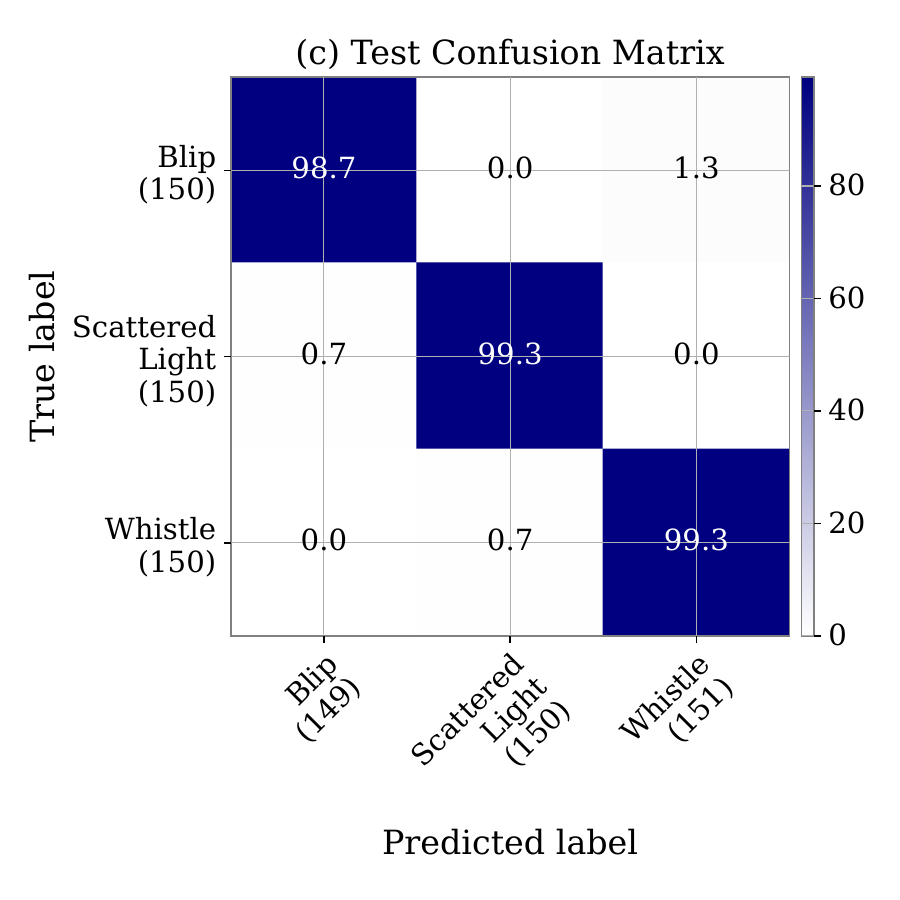}
    \end{minipage}
    
    %TC:ignore
    \caption{\justifying\textbf{(a)} Training and validation learning curves of ResNet based architecture, shown with the time axis (shift and dilation) and energy scale (clipping) augmentation.
    %shown both with and without targeted augmentation along the time axis (dilation) and energy scale (clipping).
    Augmentation improves convergence stability and enhances sensitivity to low-SNR transient features; \textbf{(b)} Confusion matrix computed on the validation set, introducing the principal glitch morphologies used in this study and illustrating the encoder's ability to separate them prior to saliency extraction. \textbf{(c)} Confusion matrix computed on a test set, following the full inference process with a second saliency-guided inference.}
    %TC:endignore
    
    \label{fig:learning_results}
\end{figure*}

We first evaluate the training performance and classification efficiency of the model. The network was trained for 200 epochs using a dataset of 4500 glitches equally split among classes. 
\review{This work is first of all a proof of concept on a glitch extraction method, therefore no optimization to accelerate the subtraction has been done.
%, nevertheless knowing that the CWT and DWT are based on Fast Fourier Transform that has a $N\log(N)$ complexity and the model is based on \texttt{pyTorch} matrices, an optimized subtraction time could tend to be of the order of a few seconds.
The model uses \texttt{pyTorch} \cite{pytorch}.}
% Add the CPU and GPU here + order of magnitude time for training and inference
\review{It requires the order of 4 hours (75 seconds per epoch in average over 200 epochs) to train.
After the training, one classification needs less than $40\,{\rm ms}$ and about 40 seconds for the segmentation-subtraction process. We used one GPU (NVIDIA RTX 4500 Ada Generation) for training and inference. The CPUs are Intel(R) Xeon(R) Gold 6526Y processors, and there are 32 of them.}

The corresponding learning curves are shown in Fig.~\ref{fig:learning_results}a.
The ground-truth labels were obtained from GravitySpy classifications. To minimize the impact of mislabeled samples and reduce entropy in the training set, only glitches with a GravitySpy confidence level above 95\% were retained.
It happens that multiple glitches occurs in a thin time windows \review{smaller than 4 seconds}. This kind of behavior would impact the training, especially when the multiple glitches have belong to different classes. Based on the omicron trigger time, such multi-glitches event are excluded from the dataset.
This selected dataset used in the training is referred as the "golden dataset" in Fig.~\ref{fig:workflow}.

The lower panels of Fig.~\ref{fig:learning_results}b and Fig.~\ref{fig:learning_results}c present the confusion matrices obtained, respectively, from the validation dataset at the 190\textsuperscript{th} epoch and from a test set independent from the 4500 previous events containing 450 glitches. \review{This test set follows the full procedure with multiple inferences as described above in the subsubsection \ref{subsubsec:inference}, the second inference being on an input image whose span is guided by the CAM.}
Although the classification accuracy during training reached 97.6\%, the complete inference phase must be evaluated independently, since the method relies on two successive evaluations, including a saliency-guided localization of the glitch using the CAM. \review{Therefore with the test set,} the resulting confusion matrices demonstrate classification accuracies ranging from 98.7\% to 99.3\% across the three classes, corresponding to an overall accuracy of 99.1\%. These results confirm the robustness and reliability of the proposed classification framework under realistic inference conditions.

Since the second classification stage is performed on a cropped spectrogram centered on the glitch and whose optimal time window is determined from the CAM localization, the resulting accuracy reflects not only the classifier performance itself, but also the ability of the CAM to preserve the characteristic glitch features required for reliable classification.
These results therefore provide a first indication that the CAM successfully identifies the relevant time--frequency structures that will subsequently be used for mask construction in the subtraction procedure.

\subsection{Extraction of a typical Whistle}

% DESCRIPTION:
%   Demonstration of empirical reconstruction using DWT masking driven by saliency.
%   Showcases both glitch-only and glitch-cleaned outputs.
%
% FIG.3: (a) Original spectrogram with saliency/outline overlay (in red)
%        (b) DWT and inverse-DWT representation with masked coefficients (overlay in red)
\begin{figure*}[ht]
    \centering
    \includestandalone[width=\linewidth]{tkz-figure/one-event-workflow}
    
    \caption{\justifying This example illustrates the ability of the method to transfer time-frequency saliency into an invertible multi-resolution scale domain. \textbf{(a)} The input CWT image. In order to highlight the glitch the spectrogram is clipped at 10. The mother wavelet used is a complex Morlet with the parameter $0.25$ and $8$ with a L2-norm.\textbf{(b)} The CAM corresponding to the inference on the input image. The red dashed box is the coarse saliency bounding box of width $\Delta t$ and height $\Delta f$. \textbf{(c)} The computation of the mask. The two plots included show an example of distributions of CAM pixel's inside (red) or outside (blue) use to select the threshold of the binarization required to build the mask shown on the left plot. \textbf{(d)} The wavelet coefficients in the DWT space. In black the transformed input signal and in red the signal after application of the mask. \textbf{(e)} The CWT spectrogram of the output signal. The same parameters and treatment are applied to the output than the input. \label{fig:wavelet_reconstruction}}
    %TC:endignore
    
\end{figure*}

% [] Show the CWT before, CAM, Mask, DWT (superimposed before-after) CWT after in one figure

Accurate glitch classification and localization constitute the first step toward signal subtraction. Figure~\ref{fig:wavelet_reconstruction} illustrates the complete reconstruction pipeline on a representative example. Starting from the original CWT spectrogram (Fig.~\ref{fig:wavelet_reconstruction}a), the combined CAM highlights the time--frequency region associated with the glitch (Fig.~\ref{fig:wavelet_reconstruction}b). The saliency-guided localization refinement further restricts this region, yielding a more precise estimate of the glitch extent in both time and frequency.

The resulting mask, derived from the CAM distribution analysis shown in Fig.~\ref{fig:wavelet_reconstruction}c, is transferred into the DWT domain. Applying this mask selectively preserves the wavelet coefficients associated with the localized transient while suppressing the surrounding background. The reconstructed signal, shown in red in Fig.~\ref{fig:wavelet_reconstruction}d, therefore contains primarily the glitch contribution.

The effectiveness of the procedure is confirmed by the CWT of the reconstructed time series displayed in Fig.~\ref{fig:wavelet_reconstruction}e. The transient structure remains clearly visible, whereas a large fraction of the surrounding noise has been removed. This example demonstrates that the proposed approach can successfully transfer saliency information from the CWT representation to an invertible multiresolution wavelet domain, enabling the extraction of a localized glitch signal in the time domain.

\subsection{Performance across glitch morphologies \label{subsec:umap}}

% FIG.5: UMAP embedding with and without (dt,df) constraints.
\begin{figure*}[ht]
    \centering
    \begin{minipage}{0.49\linewidth}
        \centering
        %\fbox{\rule{0pt}{200pt}\rule{0.90\linewidth}{0pt}}
        \includegraphics[width=\linewidth]{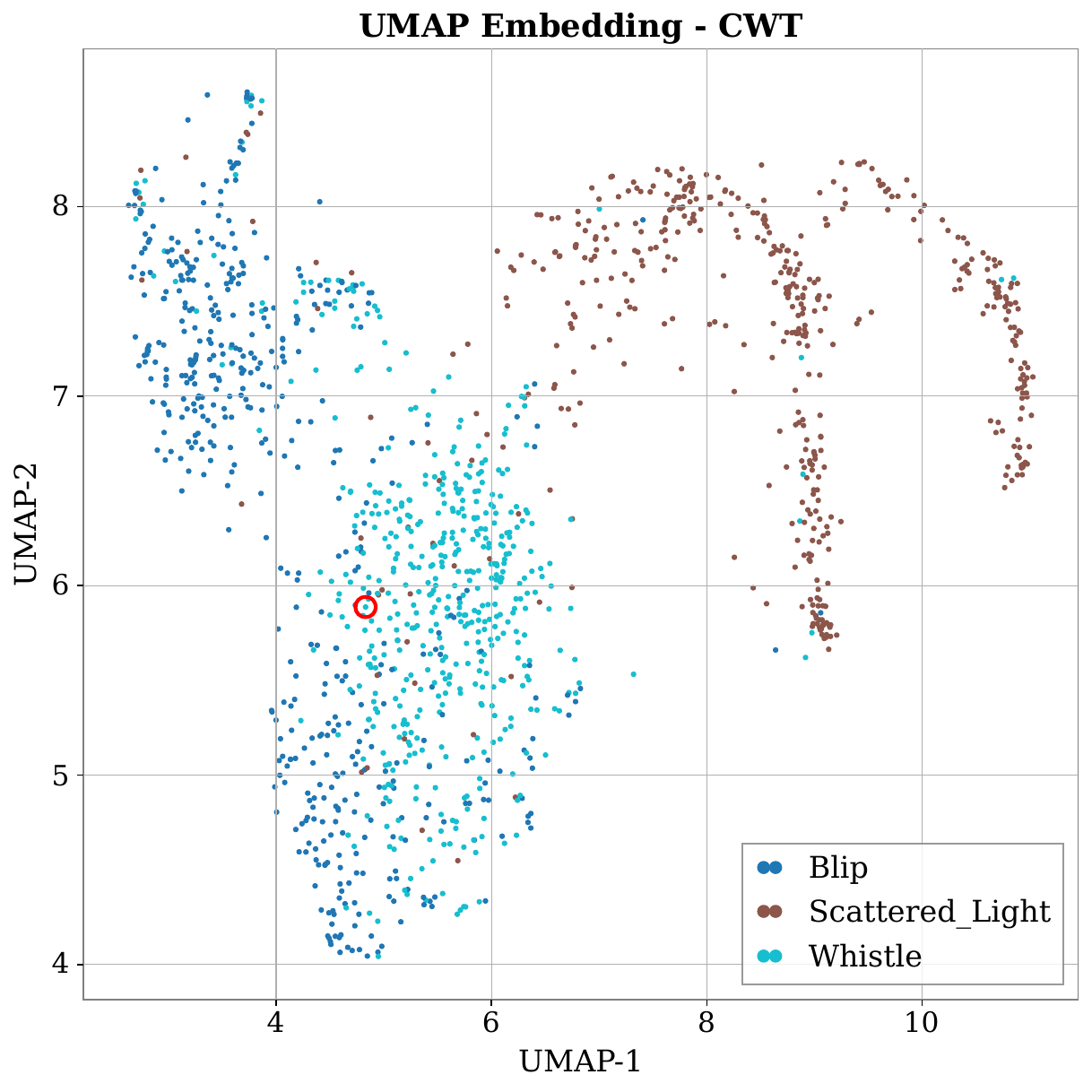}
        \subcaption{UMAP Embedding - CWT}
    \end{minipage}
    \hfill
    \begin{minipage}{0.49\linewidth}
        \centering
        %\fbox{\rule{0pt}{200pt}\rule{0.90\linewidth}{0pt}}
        \includegraphics[width=\linewidth]{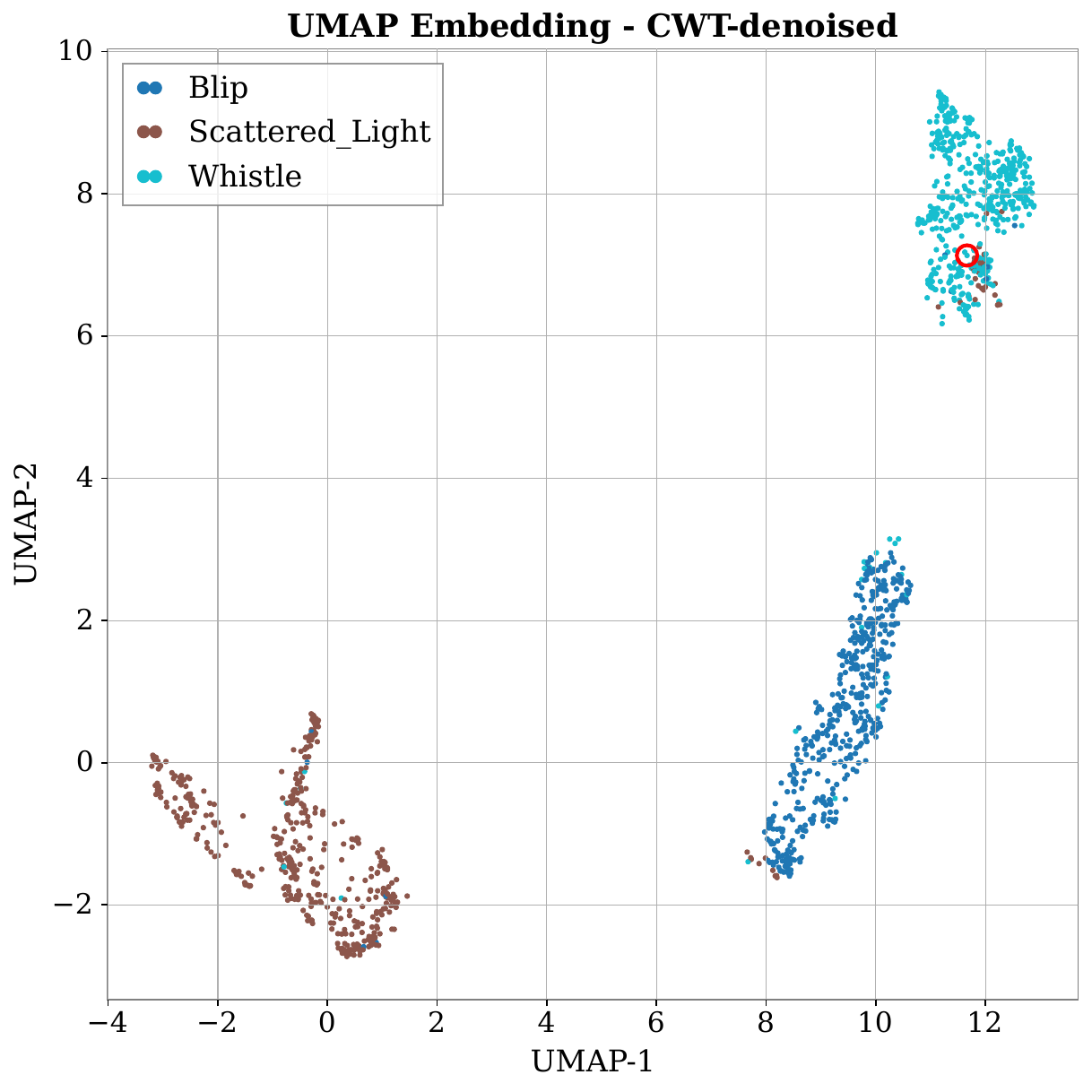}
        \subcaption{UMAP Embedding - CWT denoised}
    \end{minipage}

    %TC:ignore
    \caption{\justifying UMAP latent-space representations in two dimensions based on the learned feature embeddings. The dot colors are associated to the classification of the model used. On \textbf{(a)}, embedding obtained from CWT event spectrograms before applying the denoising. On \textbf{(b)}, embedding obtained form CWT event spectrograms after the denoising to keep the glitch and remove the surrounding gaussian noise.
    The red circle allows to show the displacement of an event.}
    %TC:endignore

    \label{fig:umap}
\end{figure*}

% DESCRIPTION:
%   Demonstrates generalization across glitch families such as whistles and scattered-light cases.
%   Emphasizes fidelity and stability of the extraction.

%In this work, we focus on three representative glitch topologies, for which a common source of confusion, particularly in the absence of data augmentation, would be the compression of \textit{whistle} patterns into narrow \textit{blip} structures depending on the time window. %In such cases, the learning architecture may fail to disentangle the underlying morphology.%, leading to incorrect feature extraction.

The previous subsection illustrated the performance of the subtraction procedure on an individual event. To assess its overall effectiveness, the analysis must be extended to a larger population of glitches and evaluated using aggregate statistical measures.

% To keep
In Fig.~\ref{fig:umap}, we evaluated the original signal \review{$s(t)$} and the reconstructed representations using CWT spectrogram via UMAP.
The embedding takes the 4 seconds CWT spectrograms without clipping, which are normalized between 0 and 1, and arbitrary parameters.
This comparison confirms that the extracted glitch preserves its morphology, bandwidth, and temporal evolution.
The CWT representation is particularly effective for visualizing such isolated glitches, in contrast to the Q-transform, for which the removal of the noise background leads to an ill-defined SNR and therefore precludes its use as a simple quantitative metric.

In these UMAP projections shown in Fig.~\ref{fig:umap}, we produced embeddings for the noisy configuration and the denoised configuration. Both of the embedding share the same events, 500 glitch event for each one of the three classes.
In the noisy configuration shown in Fig.~\ref{fig:umap}a, the embedding tends to organize into three families of glitch without a clear separation of the clusters and a considerable confusion especially between \textit{whistles} and \textit{blips}. 
On the other hand, in the denoised configuration shown in Fig.~\ref{fig:umap}b, the embedding gives a clear separation of the three clusters.
\review{Knowing the treated glitches randomly taken from the observing runs (O1, O2, O3), one can observe the differences among a same class through the runs don't prevent neither the subtraction nor the clustering performed by UMAP.}
Nevertheless, a residual confusion remains in the denoised projection. The two possible sources of confusion are either a misclassified event by the model but correctly embedded by UMAP, or the other way around, a good classification by the model but misplaced by UMAP. 
Looking down to the CWT spectra of the confused events, it appears the labels match with the glitch classes which indicate the UMAP embedding role in the confusion.

\subsection{Fainted and multi-glitches cases}

%
% FIG.6: (a) Leo-type glitch showing neighbors/sibling complexity
%        (b) Extracted feature plot

As discussed during the training analysis, multiple glitches may occur within a short time interval.
The time windows considered in this study (ranging from 0.5 to 4 seconds) generally allow unrelated glitches to be excluded from the analysis. This time-axis augmentation therefore provides an effective solution in most cases. However, when two glitches occur in close temporal proximity, isolating a single glitch within the selected time window becomes impossible.

\begin{figure*}[ht]
    \centering
    \includegraphics[width=\textwidth]{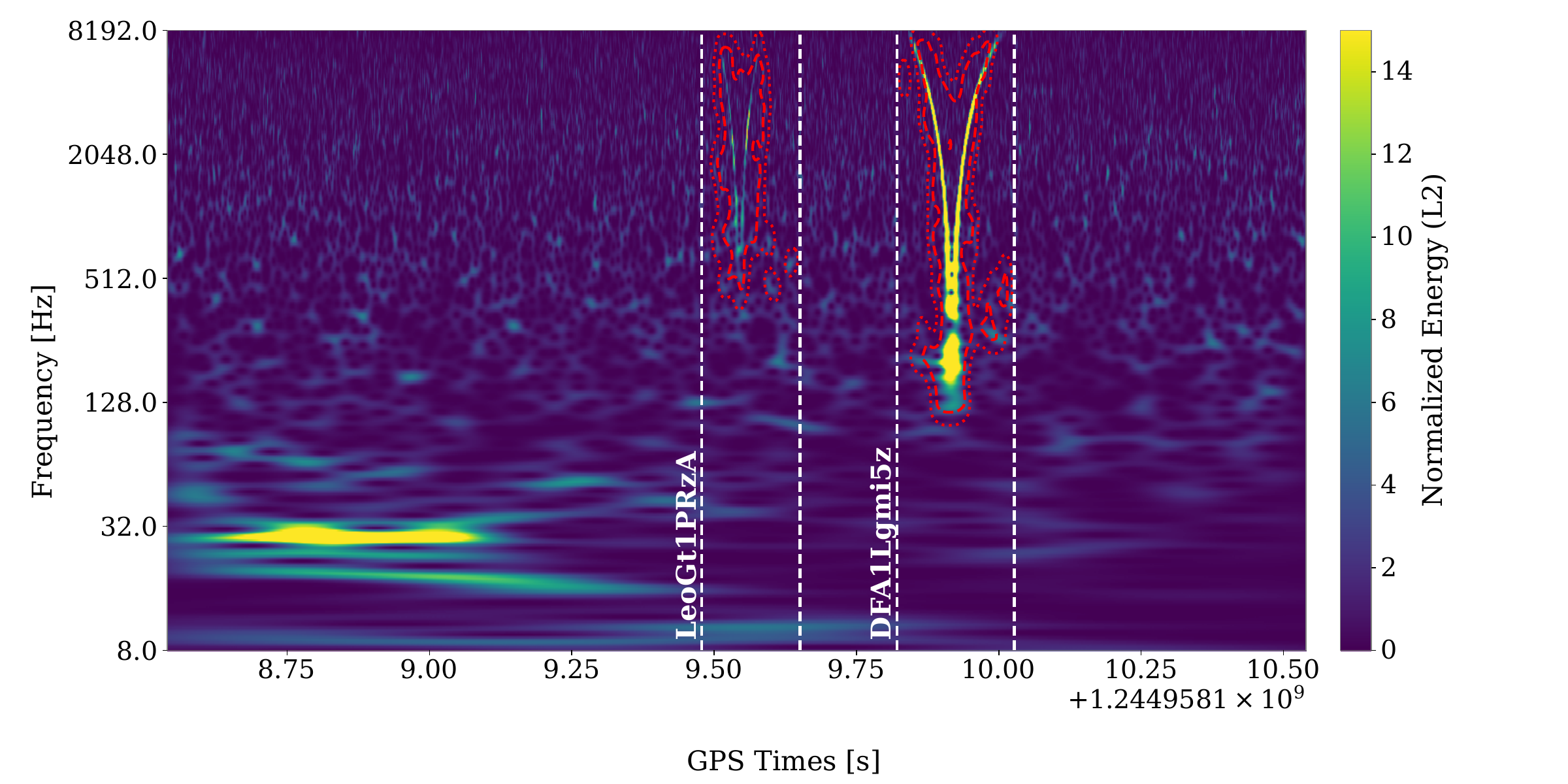}

    %TC:ignore
    \caption{\justifying CWT representation of a low-SNR Whistle-type transient \texttt{LeoGt1PRzA} (left Whistle), exhibiting multiple neighbouring sibling components (\textit{e.g.}, \texttt{DFA1Lgmi5z} (right Whistle), \texttt{6GooD7MdgA}, \texttt{jTITTOjjn4}) occurring in close temporal and frequency proximity.
    The two glitches \texttt{LeoGt1PRzA} and \texttt{DFA1Lgmi5z} are shown in surounded by the white dashed vertical lines from the saliency-guided time location. 
    The red lines are contours of the mask respectively at value 0.9 and 0.5 for the dashed and the dotted lines.
    %Feature activation prior to wavelet-domain masking, illustrating the model's ability to isolate the faint structure in the presence of significantly higher-SNR transients and under complex overlapping conditions.}
    }
    %TC:endignore
    \label{fig:leo_complexity}
\end{figure*}

To address this issue, the input CWT spectrum is processed twice, each pass targeting a different glitch subtraction. \review{As shown in Fig.~\ref{fig:leo_complexity}}, during the first pass, the \texttt{DFA1Lgmi5z} glitch is subtracted ; owing to its higher SNR, this glitch dominates the model attention. The resulting subtraction is then used to mask this glitch before applying a second pass to subtract the \texttt{LeoGt1PRzA} glitch.
For both glitches shown in Fig.~\ref{fig:leo_complexity}, the white dashed vertical lines indicate the times derived from the CAM-localized bounding boxes.
In configurations where multiple glitches cannot be isolated through time-window selection, this iterative subtraction procedure is performed automatically.

This example demonstrates the ability of the method to isolate and subtract nearby glitches independently. In particular, the subtraction of the \texttt{LeoGt1PRzA} glitch also illustrates the sensitivity of the method to low-amplitude glitches with amplitudes close to the noise level.

\section{Discussions}
% %%%%%%%%%%%%%%%%%%%%%%%%%%%%%%%%%%%%%% %
% This is the discussions section (III). %
% %%%%%%%%%%%%%%%%%%%%%%%%%%%%%%%%%%%%%% % 

% 1. High-level significance of the results:
% KEYPOINTS:
% [X] Importance of isolating transient noise directly in the time domain.
% [X] Novelty of combining saliency with invertible wavelet reconstruction.
The results presented in this work demonstrate that time-frequency saliency, when combined with an invertible multiresolution representation and appropriate training augmentation, provides an effective mechanism for isolating %\review{\sout{a wide variety of}}
transient noises. 
This capability is highly relevant for GW detector characterization, where the ability to extract, suppress, or reconstruct specific glitch families is essential for data-quality assessment and for the stability of astrophysical searches. The proposed framework introduces a practical bridge between non-invertible saliency techniques, traditionally use as explainability rationale\cite{Koyama:ClassificationRationale:2024} for classification, and a fully invertible wavelet reconstruction pipeline. This combination enables the production of glitch-only and glitch-cleaned strain time series without relying on surrogate modeling.\\

% 2. Why the segmentation framework works:
% KEYPOINTS:
% [X] Interpretability: saliency maps highlight physically meaningful structures.
% [X] Invertibility of DWT enables exact reconstruction, unlike Q-transform/CWT.
% [X] Strengths compared to purely supervised classifiers (e.g., Gravity Spy).
The effectiveness of the proposed approach is the interpretability and localization provided by the saliency maps. By identifying compact regions in the time-frequency plane to which the model assigns the highest discriminative relevance, the saliency maps highlight structures that often correspond to physically meaningful components of the glitch.
When mapped to the DWT domain, this information is translated into a sparse set of coefficients at specific scales and temporal locations. Because the DWT is exactly invertible, the selected coefficients can be recombined to yield a faithful reconstruction of the transient—an outcome that is not achievable with traditional SNR-weighted Q-transform or CWT representations.
This capability extends beyond purely supervised classification strategies, which provide labels or embeddings but do not enable waveform extraction or the automated generation of glitch-free data.

% 3. Implications for GW detector characterization:
% KEYPOINTS:
% [X] Controlled extraction of glitches for generative modeling.
% [X] Glitch-free reconstruction for background estimation.
% [X] Applicability to data-quality studies, calibration tests, and pipeline validation.
From the perspective of detector characterization, the ability to selectively retain or suppress specific components of the strain has several direct implications.
First, the extracted glitches provide isolated instances of well-defined morphology that can serve as inputs for generative modeling.
Second, the glitch-free reconstructions provide an alternative representation of the underlying strain that can be used to improve background estimates in matched-filter and excess-power analyses.
Finally, the masking procedure allows for controlled examination of localized features, making it applicable to data-quality investigations, calibration assessments, and validation of low-latency and offline search pipelines, particularly in cases where transient noise may influence trigger conditions or parameter estimation.\\

% 4. Limitations:
% KEYPOINTS:
% [X] Sensitivity to extremely low-SNR transients.
% [X] Resolution constraints from wavelet decomposition.
% [X] Dependence on the quality of saliency localization.
Despite these advantages, our method has certain limitations that may naturally suggest directions for future work.
First, the framework is not designed to recover or disentangle transient noise when it is temporally or morphologically correlated with an actual gravitational-wave signal.
In such cases as GW170817\cite{LIGO:GW170817:2017}, glitches overlapping with events, accurate separation would require a dedicated Bayesian inference framework capable of jointly modeling both the astrophysical waveform and the noise transient.
Second, the spatial resolution of the reconstruction is constrained by the properties of the chosen wavelet family and by the Gabor-Heisenberg uncertainty principle, which impose fixed relationships between temporal and frequency resolution across scales. % Possible solution with the synchrosqueezing (on top of the CWT)
Finally, the method mainly relies on the refinement of the saliency mask through postprocessing: although the raw saliency can be diffuse or imperfect, the postprocessing step stabilizes the localization and ensures that only the relevant structures are projected onto the wavelet domain.\\% Possible solution by combinings CAMs or choosing a different CAM family, suiting more to our data

% 5. Future directions:
% KEYPOINTS:
% [X] More glitch classes.
% [X] Application to O3GK and O4 datasets.
% [X] Integration with scalable detection architectures (e.g., YOLO-style real-time models).
% [X] Use of RGB structure with a modified approach to refine the detectability
Looking ahead about future research work, several directions appear promising. Applying the framework to full O1-O3 datasets, O3GK, and O4 datasets will enable a systematic assessment of its utility in large-scale detector-characterization studies, with potentially useful implications for the design of future gravitational wave interferometers. \review{Indeed, characterizing and studying glitches class by class may allow one to have a better understanding of their origin and thus fine tune the design of detectors in order to minimize or avoid some kind of glitches.}

Integration with more performant and scalable detection architectures, such as YOLO-based real-time segmentation, could extend the method to continuous, low-latency monitoring while simplifying the computation of saliency maps.
Some works on glitch segmentation with YOLO have already been performed by \cite{yolo-glitch}.
In this context, the contours produced by our current framework could directly serve as ground truth during the initial training of YOLO models.
%Alternatively, utilizing the RGB input layer of the ResNet model’s, together with a modified reconstruction strategy, may improve sensitivity to weak or overlapping transients by incorporating three complementary views of the same spectrogram, for instance, different CWT-based normalizations (L1, L2) with in addition the Synchrosqueezed signal \cite{synchrosqueeze_daubechies}.
Taken together, these developments could establish a new class of hybrid models that combine interpretability, invertibility, and high detection performance for the next generation of gravitational-wave data analysis.
\section{Conclusion
    %\wordcount{sections/5_conclusion.tex}{1000}
    }
% %%%%%%%%%% %
% Conclusion %
% %%%%%%%%%% %

% Short, reminds the contribution, main results, prospects
% [X] Contibution : probleme, method, novelties
% [X] Main results : Extraction, invertability, interpretability, utility for GW studies
% [X] Prospects : O4/O5, low-latency, scalable, future detector characterisation

In this work, we introduced a saliency-guided framework for the identification and subtraction of transient glitches in gravitational-wave strain data. By combining CAM localisation with an invertible discrete wavelet transform representation, the proposed method enables the extraction and subtraction of transient noise components directly in the time domain while preserving the strain data.

The results demonstrate that saliency information derived from a supervised classifier can be effectively translated into sparse wavelet-domain masks that retain the characteristic time--frequency morphology of several glitch families. Unlike conventional non-invertible time--frequency representations, the proposed approach provides direct access to reconstructed glitch-only and glitch-cleaned strain time series, thereby extending the role of saliency methods beyond interpretability alone.

Beyond classification, this framework provides a practical tool for detector-characterisation studies, data-quality investigations, and future low-latency analysis strategies. Its compatibility with scalable segmentation architectures and large observational datasets suggests that invertible saliency-based approaches could become valuable components of next-generation gravitational-wave data-analysis pipelines.\\

\begin{acknowledgements}

This research was supported in part by the Japan Society for the Promotion of Science (JSPS) Grant-in-Aid for Fellows [No. P21726 (M.\ Meyer-Conde)], the JSPS Postdoctoral Fellowships for Research in Japan [No. P25701 (C.\ Alléné)] and the JSPS Grant-in-Aid for Scientific Research [Nos. 23H01176, 23K25872 and 23K22499] (H.\ Takahashi)]. This research was supported by the Joint Research Program of the Institute for Cosmic Ray Research, University of Tokyo, and Tokyo City University Prioritized Studies and Research Equipment Program.\\

This research has also made use of data or software obtained from the Gravitational Wave Open Science Center (gwosc.org), a service of the LIGO Scientific Collaboration, the Virgo Collaboration, and KAGRA. This material is based upon work supported by NSF's LIGO Laboratory which is a major facility fully funded by the National Science Foundation, as well as the Science and Technology Facilities Council (STFC) of the United Kingdom, the Max-Planck-Society (MPS), and the State of Niedersachsen/Germany for support of the construction of Advanced LIGO and construction and operation of the GEO600 detector. Additional support for Advanced LIGO was provided by the Australian Research Council. Virgo is funded, through the European Gravitational Observatory (EGO), by the French Centre National de Recherche Scientifique (CNRS), the Italian Istituto Nazionale di Fisica Nucleare (INFN) and the Dutch Nikhef, with contributions by institutions from Belgium, Germany, Greece, Hungary, Ireland, Japan, Monaco, Poland, Portugal, Spain. KAGRA is supported by Ministry of Education, Culture, Sports, Science and Technology (MEXT), Japan Society for the Promotion of Science (JSPS) in Japan; National Research Foundation (NRF) and Ministry of Science and ICT (MSIT) in Korea; Academia Sinica (AS) and National Science and Technology Council (NSTC) in Taiwan.
\end{acknowledgements}

\appendix
\section{Complex Morlet parameters}
\begin{figure*}
    \centering
    \includegraphics[width=\linewidth]{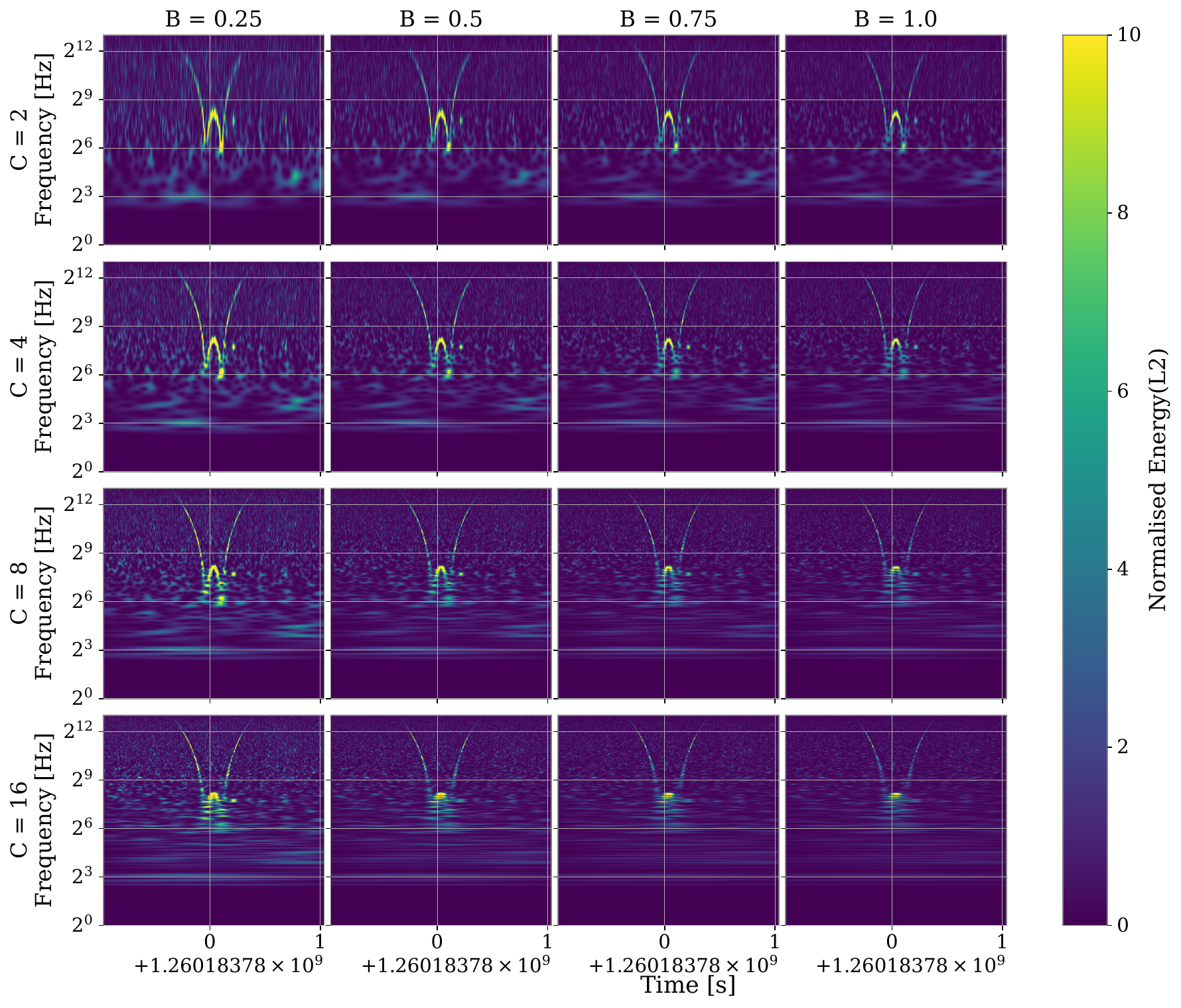}

    %TC:ignore
    \caption{\justifying CWT spectrogram of a Whistle glitch. The normalization is L2 and the mother wavelet is a complex morlet.}
    %TC:endignore
    \label{fig:cmor_param}
\end{figure*}

In the presentation of the continuous wavelet transform in sub-subsection \ref{subsec:cwt} a choice is made on the $B$ and $C$ parameters. In Fig.~\ref{fig:cmor_param} a set of spectrograms with different $B$ and $C$ values is displayed. It shows how the energy gets diffused with larger $B$, indeed this parameter is proportional to the standard deviation of the gaussian component of the mother wavelet. It also shows how the central frequency related to the $C$ parameters modify the time and frequency spreading. Both of these parameters impact the time-frequency localization of the energy.

\section{Class Activation Map classes}
\begin{figure*}
    \centering
    \includegraphics[width=0.8\linewidth]{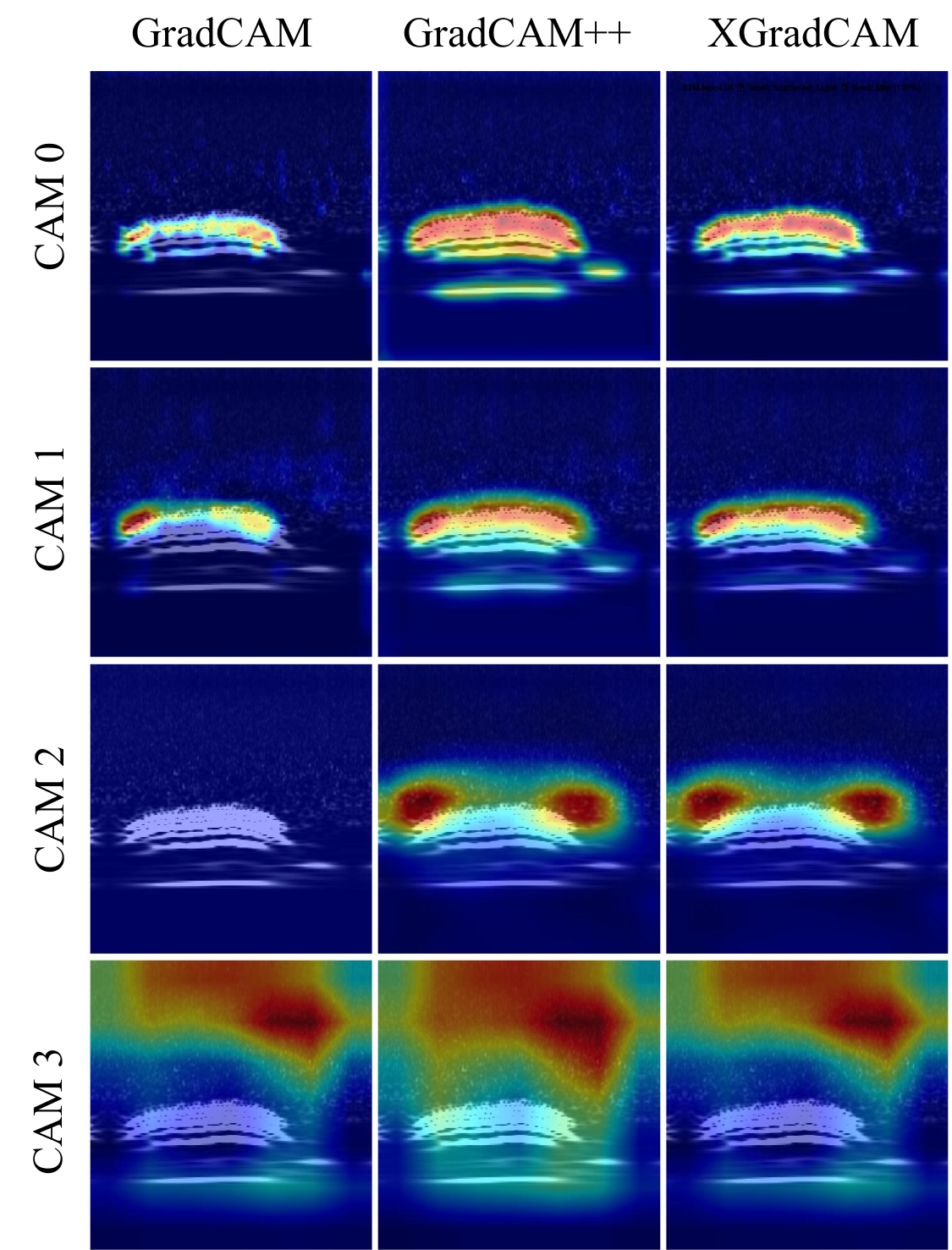}

    %TC:ignore
    \caption{\justifying Example of GradCAM, GradCAM++ and XGradCAM use over the last layers of the 4 bottlenecks.}
    %TC:endignore
    \label{fig:cam_compare}
\end{figure*}

As discussed in subsubsection \ref{subsubsec:cam}, the GradCAM++ algorithm allows to highlight the spatially spread component of the classified instance (the glitch in our case). this behavior is well shown in Fig.~\ref{fig:cam_compare}, especially for the two first layers displayed when compared with the original GradCAM or XGradCAM.
Moreover, the seperated components of the glitch shown are highlighted by GradCAM++. This glitch is a salient example about why the GradCAM++ has been chosen over GradAM or XGradCAM.
Nevertheless, XGradCAM, as is shown in the first three layers, has a better conservation of the contribution of the characteristics. Therefore, this justifies the use of XGradCAM over GradCAM when GradCAM++ fails to provide a map.
According to the test set of 1500 glitches presented above, the GradCAM++ algorithm didn't failed to produce a map leading to a proportion of XGradCAM usage lower than $0.07\%$. In any case, this fallback was kept as a precaution.

\section{Glitch proportion through observing run}
\begin{table}[h!]
    \centering
    \begin{subtable}{\columnwidth}
    \centering
    \begin{tabular}{c|ccc|c}
       & Blip & Scattered Light & Whistle & Total\\\hline
      O1 & 80 (5.3\%) & 38 (2.5\%) & 0 (0\%) & 118 (2.6\%) \\ 
      O2 & 821 (54.7\%) & 207 (15.8\%) & 229 (15.3\%) & 1257 (27.9\%) \\
      O3a& 278 (18.5\%) & 205 (13.7\%) & 564 (37.6\%) & 1047 (23.3\%)\\
      O3b& 321 (21.4\%) & 1050 (70.0\%) & 707 (47.1\%) & 2078 (46.2\%)\\\hline
      Total & 1500 & 1500 & 1500 & 4500
    \end{tabular}
    \caption{Table of the glitch distribution among observing runs for the training dataset.}
    \end{subtable}
    \begin{subtable}{\columnwidth}
    \centering
    \begin{tabular}{c|ccc|c}
       & Blip & Scattered Light & Whistle & Total\\\hline
    O1 & 40 (8.1\%) & 34 (6.9\%) & 31 (6.1\%) & 105 (7.0\%)\\
    O2 & 247 (49.9\%) & 78 (15.8\%) & 45 (8.8\%) & 370 (24.7\%)\\
    O3a& 96 (19.4\%) & 53 (10.7\%) & 207 (40.7\%) & 356 (23.8\%)\\
    O3b& 112 (22.6\%) & 329 (66.6\%)& 223 (44.4\%) & 667 (44.5\%)\\\hline
    Total&495 & 494 & 509 & 1498
    \end{tabular}
    \caption{Table of the glitch distribution among observing runs for the test dataset.}
    \end{subtable}
    \caption{Table of the glitch distributions among the observing runs for both the training and test dataset.}
    \label{tab:run_prop}
\end{table}

\begin{table}[h!]
    \centering
    \begin{subtable}{\columnwidth}
    \centering
    \begin{tabular}{c|ccc}
      & Blip vs S. Light & Blip vs Whistle & S. Light vs Whistle\\\hline
      $R^2$& 0.692&0.787&0.790
    \end{tabular}
    \caption{Table of coefficient of determination between glitch classes.}
    \end{subtable}
    \begin{subtable}{\columnwidth}
    \centering
    \begin{tabular}{c|ccc}
      Class & Blip & Scattered Light & Whistle\\\hline
      $R^2$&$7.59\times10^{-2}$&$4.74\times10^{-2}$&$3.00\times10^{-2}$
    \end{tabular}
    \caption{Table of coefficient of determination between the O1/2 runs and the O3 run.}
    \end{subtable}
    \caption{Tables of coefficient of determination. The coefficient is computed on the 2D UMAP representation of the CWT spectrogram containing only the subtracted glitch. The seed of the UMAP is 0 and the parameters are $\texttt{min-dist}=0.01$ and $\texttt{n-neighbor}=15$.}
    \label{tab:R2}
\end{table}
During the training and the test the glitches has been equally distributed among the classes: a third of Whistle, a third of Blip and a third of Scattered Light. However, the distribution among the observing run are as shown in Tab.~\ref{tab:run_prop}. 

Given a 2D UMAP representation of the CWT spectrogram containing only the subtracted glitch, one can, for a given seed (here 0), compute the coefficient of determination between two groups A and B:
\begin{equation}
    R_{A, B}^2=\frac{SS^{A,B}_{\rm between}}{SS^{A,B}_{\rm between}+SS^{A,B}_{\rm within}}
\end{equation}
with the sum of squares between-groups and within-groups:
\begin{equation}
\begin{aligned} 
SS^{A,B}_{\rm between} = \frac{n_An_B}{n_A+n_B}||\bar{x}_A-\bar{x}_B||^2\\
SS^{A,B}_{\rm within} = \sum_{i\in A}||x_i-\bar{x}_A||^2 + \sum_{i\in B}||x_i-\bar{x}_B||^2
\end{aligned}
\end{equation}
where A and B are two populations among the dataset, $x$ denote the position in the 2D UMAP space and the bar indicates the centroid position of a population.
For a UMAP representation with the parameters $\texttt{min-dist}=0.01$ and $\texttt{n-neighbor}=15$, we obtain the coefficients of the Tables~\ref{tab:R2}.

These tables show the clear separation one can already observe in Fig. \ref{fig:umap} with $R^2$ values of the order of 0.7. and a negligible separation between the glitches of O1/2 and those of O3 ($R^2\sim \mathcal{O}(10^{-2})$) compared to the spearation between classes. This sustains the observations of the Fig. \ref{fig:umap} showing three compact clusters.

\bibliographystyle{apsrev4-2}
\bibliography{main}

\end{document}